\documentclass[nofootinbib,amsmath,twocolumn,notitlepage,preprintnumbers,floatfix]{revtex4-1}
\usepackage{multirow}
\usepackage{amssymb, esvect, amsmath, graphicx, latexsym, amsthm, slashed, eso-pic}
\usepackage{xcolor}
\usepackage{amsmath,amsthm,amsfonts,amscd,amssymb} 
\usepackage{eucal}
\usepackage{here}
\usepackage{braket}
\usepackage{slashed}
\usepackage{hyperref}
\usepackage{graphicx}
\usepackage{bm}
\usepackage{siunitx}
\usepackage{xcolor}
\usepackage{afterpage}
\usepackage{changepage}
\usepackage{soul}
\usepackage{float}
\usepackage{mathrsfs}
\usepackage{here}
\usepackage{float}
\usepackage{mathrsfs}
\usepackage{wrapfig}
\usepackage{colortbl}
\usepackage{siunitx} 
\usepackage{ascmac}
\usepackage{subfiles}
\usepackage{comment}
\usepackage{slashed}

\usepackage{cleveref}
\usepackage{mathtools}
\usepackage[normalem]{ulem}
\hypersetup{
    colorlinks=true,
    linkcolor=blue,
    citecolor=magenta,      
    urlcolor=cyan}

\usepackage{lipsum}

\def\atag{\refstepcounter{equation}\tag{\thesection\arabic{equation}}}
\newcounter{num}

\usepackage{hyperref}

\usepackage[english]{babel}

\begin{document}

\title{Old neutron stars as a new probe of relic neutrinos and sterile neutrino dark matter}

\author{Saurav Das$^a$}
\thanks{s.das@wustl.edu}
\author{P.~S.~Bhupal Dev$^a$}
\thanks{bdev@wustl.edu }
\author{Takuya Okawa$^{a,b}$}
\thanks{o.takuya@wustl.edu }

\author{Amarjit Soni$^c$}
\thanks{adlersoni@gmail.com}

\medskip

\affiliation{$^a$ Department of Physics and McDonnell Center for the Space Sciences, Washington University, Saint Louis, MO 63130, USA}
\affiliation{$^b$ Theoretical Physics Department, Fermi National Accelerator Laboratory, Batavia, IL 60510, USA}
\affiliation{$^c$ High Energy Theory Group, Physics Department, Brookhaven National Laboratory, Upton, NY 11973, USA  }

\date{\today}

\begin{abstract}
We study the kinetic cooling (heating) of old neutron stars due to coherent scattering with relic neutrinos (sterile neutrino dark matter)  via Standard Model neutral-current interactions. We take into account several important physical effects, such as coherent enhancement, gravitational clustering, neutron degeneracy, Pauli blocking and weak potential. We find that the anomalous cooling of neutron stars due to relic neutrino scattering is difficult to observe. However, the anomalous heating of neutron stars due to coherent scattering with keV-scale sterile neutrino dark matter 
may be observed by current and future telescopes operating in the optical to near-infrared frequency band, such as the James Webb Space Telescope (JWST), which would probe hitherto unexplored parameter space in the sterile neutrino mass-mixing plane. 
\end{abstract}

\maketitle

\section{Introduction}
Cosmic Neutrino Background (C$\nu$B) is a robust prediction of the standard cosmological model of the Universe~\cite{Weinberg:1962zza}. However, the existence of C$\nu$B has only been indirectly inferred from big bang nucleosynthesis (BBN)~\cite{Pitrou:2018cgg}, cosmic microwave background (CMB)~\cite{Planck:2018vyg}, and large-scale structure (LSS)~\cite{DESI:2024mwx} data; see Ref.~\cite{DiValentino:2024xsv} for a recent review. In spite of having the largest flux among all natural and laboratory sources of neutrinos~\cite{Vitagliano:2019yzm}, direct detection of C$\nu$B remains elusive~\cite{Gelmini:2004hg,Yanagisawa:2014, Vogel:2015vfa}. This is mainly due to its small kinetic energy today which results in extremely small cross-section and energy/momentum transfer in a laboratory setup~\cite{PTOLEMY:2018jst}. 

\begin{figure}[t!]
    \centering
    \includegraphics[width=0.45\textwidth]{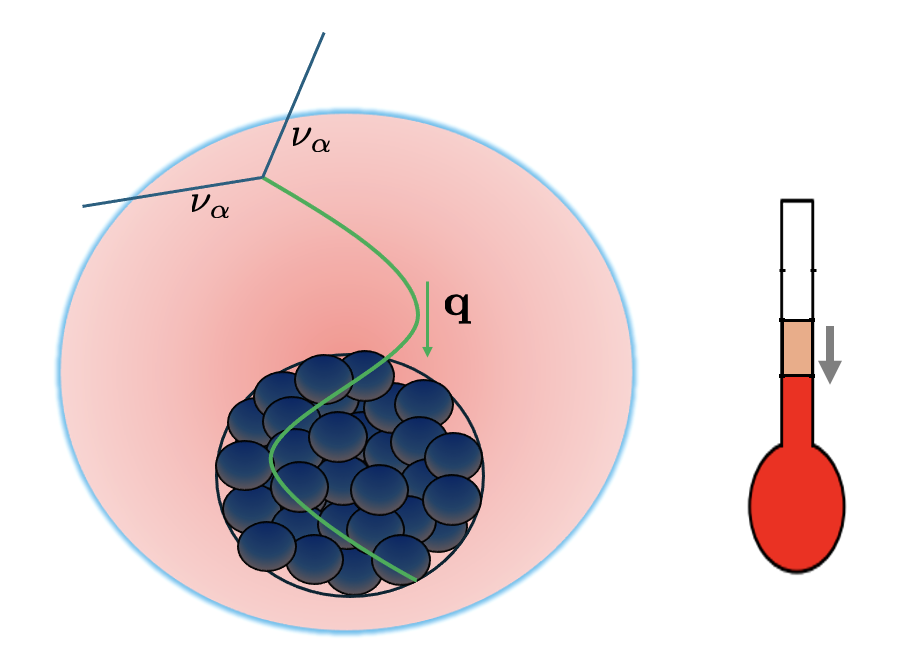}
    \caption{An artist's rendition of the NS kinetic {\it cooling} via C$\nu$B. Gravitationally captured relic neutrinos coherently scatter off neutrons  via the neutral-current interaction within a sphere of radius $\lambda_\nu=2\pi/|{\bf q}|$, allowing energy transfer from NS to C$\nu$B, thus anomalously cooling down the NS.}
    \label{fig:artist}
\end{figure}

One way to potentially overcome this difficulty is by using coherent scattering~\cite{Freedman:1973yd} where the neutrino may collectively interact with all particles within a sphere of radius equal to the de Broglie wavelength corresponding to the momentum transfer. Owing to their small momenta, relic neutrinos can receive a huge enhancement in their scattering cross-section in a dense astrophysical environment like a neutron star (NS), where they can coherently interact with a gigantic number of neutrons even in a small volume. Since C$\nu$B behaves as a cold, non-relativistic fluid, its interaction with neutrons in a hot NS medium will result in an anomalous cooling of the NS; see Fig.~\ref{fig:artist}. To the best of our knowledge, the only paper which studied this cooling effect was long ago~\cite{Dixit:1983aj}, and they found a negligible kinetic cooling rate. In this paper, we correct the old calculation by taking into account coherent scattering, gravitational clustering, neutron degeneracy and Pauli blocking effects. Although we get orders of magnitude improvement over the results in Ref.~\cite{Dixit:1983aj}, we still find the kinetic cooling rate to be subdominant compared to the standard cooling rate via black-body emission of photons, which makes the kinetic cooling effect unobservable. The presence of a large local C$\nu$B overdensity might improve the observational prospects, but as we show below, it may not be sufficient. 

We next consider the possibility of anomalous NS heating due to its interaction with heavier neutrino species with mass much larger than the NS temperature. Due to cosmological~\cite{Planck:2018vyg, DESI:2024mwx} and laboratory~\cite{Katrin:2024tvg} upper bounds on the absolute neutrino mass scale, the active neutrino mass eigenstates must be at sub-eV scale, and therefore, cannot contribute to NS heating. However, there could be additional sterile neutrino species which are much heavier. In fact, keV-scale sterile neutrinos having tiny mixing with active neutrinos can be an excellent dark matter (DM) candidate~\cite{Drewes:2016upu}. Assuming this to be the case, we find that sterile neutrino accumulation inside NS can lead to the anomalous heating of NSs (see Fig.~\ref{fig:artist2}), which is potentially detectable for a local cold old NS by optical and near-infrared telescopes such as the currently operating James Webb Space Telescope (JWST)~\cite{Gardner:2006ky}, near-future European Extremely Large Telescope (ELT)~\cite{2018sf2a.conf....3N}, and far-future Thirty Meter Telescope (TMT)~\cite{TMT:2015pvw}. Using the projected sensitivities of these instruments in terms of NS temperatures, we derive new constraints in the sterile neutrino mass-mixing plane, which are comparable to the most stringent astrophysical constraints from $X$-ray line searches.  

The rest of the paper is organized as follows. We first discuss neutron star temperature in Section~\ref{sec: NS energy loss}
followed by coherent scattering in Section~\ref{sec:enhancement factor}, then NS cooling in Section~\ref{sec: relic neutrino capture by a NS} and heating via
sterile neutrinos in Section~\ref{sec: NS_heating}. Finally we conclude with a discussion in section~\ref{sec: summary} where we emphasize 
simplifications used and some future avenues to be explored. Some calculation details and arguments in support of the assumptions made in the main text are relegated to the Appendices~\ref{app:A}-\ref{app:D}.   
\section{Neutron star temperature}
\label{sec: NS energy loss}
NSs are born hot in supernova explosions, with internal temperature $T_{\rm int}\sim 10^{11}$ K. Few seconds after their birth, they become transparent for neutrinos generated in their interiors and cool primarily via neutrino emision~\cite{Sawyer:1978qe}, although the initial cooling stage ($t\lesssim 100$ yr) is also accompanied by thermal relaxation of NS interiors. 
At later stages ($t\gtrsim 10^5$ yr) they mainly cool via thermal surface emission of photons~\cite{Yakovlev:2004yr}.
According to the minimal cooling paradigm~\cite{Page:2004fy,Page:2005fq, Ofengeim:2017cum}, old ($t>10^9$ yr) isolated NSs exhaust all their thermal and rotational energy, cooling down to temperatures below ${\cal O}(100)$ K and making them impossible to observe. 

In fact, most of the isolated NSs detected so far are young radio pulsars with $t\lesssim 10^6$ yr and surface temperature observed at infinity $T_s^{\infty}$ between $10^5$ and $10^6$ K~\cite{Yakovlev:2010ed}. Older NSs have only been detected indirectly as companions in binary systems, such as millisecond pulsars~\cite{1994Sci...264..538W, Lorimer:2008se, Lattimer:2015nhk}. The coldest known NS on record, PSR J2144-3933, located at 172 pc from Earth, has its effective surface temperature bounded at $T_s^{\infty}<42, 000$ K by the Hubble Space Telescope~\cite{Guillot:2019ugf}. Its spin-down age is $\sim 3\times 10^8$ yr, which implies that its temperature would be at most a few hundred K in the minimal cooling models~\cite{Page:2004fy, Page:2005fq, Ofengeim:2017cum}. Based on Monte-Carlo orbital simulations of the spatial distribution of galactic NSs, we expect 1-2 (100-200) such cold, old, isolated NSs lurking within 10 (50) pc~\cite{Ofek:2009wt, Sartore:2009wn, 2016RAA....16..101T}. On the observational front, the currently operating JWST~\cite{Gardner:2006ky} and future ELT~\cite{2018sf2a.conf....3N} and TMT~\cite{TMT:2015pvw} are capable of detecting black-body peak temperatures of 1300-4300 K with their optical to near-infrared imaging instruments. Therefore, it is timely to consider anomalous heating/cooling of old local NSs as a new probe of relic and sterile neutrinos. 

\section{Coherent Scattering}
\label{sec:enhancement factor}
The C$\nu$B temperature today is: $T_\nu = (4/11)^{1/3}T_\gamma \simeq 1.9\si{\ K}\simeq 1.7\times 10^{-4}$ eV, where $T_\gamma \simeq 2.7\si{\ K}$ is the CMB temperature~\cite{Planck:2018vyg}. Assuming that the neutrino rest mass $m_\nu>T_\nu$ which we know is true for at least two of the three active neutrino mass eigenstates in order to fit the observed mass-squared differences~\cite{nufit}, we can treat the C$\nu$B as a cold, non-relativistic fluid. For a relic neutrino scattering against a neutron in old NSs, since the momentum transfer $|{\bf q}| \simeq T_\mathrm{NS} \lesssim 0.1\si{\ eV}$ (where $T_{\rm NS}$ is the NS surface temperature) is very small, the corresponding de Broglie wavelength $\lambda_\nu=2\pi/|{\bf q}| \simeq 10^{-3}~{\rm cm}(0.1\si{\ eV}/|{\bf q}|)$ is much larger than the size of the nucleon, and hence, the neutrino will scatter off the entire neutron, with a total unpolarized scattering cross-section of~\cite{Formaggio:2012cpf}
\begin{align}
    \sigma_{\nu n} \simeq \frac{G_F^2}{\pi}E_\nu^2 \simeq \left(1.7\times 10^{-58}~{\rm cm}^{2}\right)\left(\frac{E_\nu}{0.1~{\rm eV}}\right)^2 \, ,
    \label{eq:crosssim}
\end{align}
where $G_F$ is the Fermi constant and $E_\nu$ is the neutrino energy. As shown in Appendix~\ref{app:A}, Eq.~\eqref{eq:crosssim} is a pretty good approximation for neutrino energies much smaller than the target mass. This cross-section is small for C$\nu$B energy, but coherent scattering with a large number of neutrons can give a significant enhancement. The cross-section for coherent elastic neutrino-nucleus scattering (CE$v$NS), up to leading order in $E_\nu/m_N$ (where $m_N$ is the nucleus mass), is given by~\cite{Formaggio:2012cpf}
\begin{align}
    \sigma_{\nu N} =\frac{G_F^2}{4 \pi}\left(Q_V^2+3 Q_A^2\right) E_{\nu}^2 \, , 
    \label{eq:coh}
\end{align}
where $Q_V = A - 2Z(1 - 2\sin^2 \theta_w)\simeq A-Z$ and $Q_A = A - 2Z$ are respectively the vector and axial charges of the nucleus, given in terms of its mass and atomic numbers $A$ and $Z$ respectively, and $\theta_w$ is the weak mixing angle.  
Note that the CE$v$NS process has already been observed by the COHERENT experiment with CsI~\cite{COHERENT:2017ipa}, Ar~\cite{COHERENT:2020iec} and Ge~\cite{Adamski:2024yqt} detectors. For an NS with mostly neutrons, we can set $A=1$ and $Z=0$, which recovers Eq.~\eqref{eq:crosssim}.  However, since the number density of neutrons ($n_n$) inside NS is huge, the coherent cross-section will be enhanced by a factor of 
\begin{align}
    N_C \simeq \frac{4}{3}\pi \lambda_\nu^3 n_n \simeq 3 \times 10^{30} \left(\frac{0.1 \si{\ eV}}{|{\bf q}|}\right)^3\left(\frac{n_n}{4 \times 10^{38}~{\rm cm}^{-3}}\right) , 
    \label{eq:3}
\end{align}
which is the number of neutrons within a sphere of radius equal to the de Broglie wavelength $\lambda_\nu=2\pi/|{\bf q}| \simeq 10^{-3}~{\rm cm}(0.1\si{\ eV}/|{\bf q}|)$.  Note that although $\lambda_\nu$ is much smaller than the typical NS radius $R_{\rm NS}\simeq 10$ km, due to the huge neutron number density, $N_C$ turns out to be a large number; see Appendix~\ref{app:B} for a more precise calculation of $N_C$.
\section{NS Cooling by C$\nu$B} 
\label{sec: relic neutrino capture by a NS}
To calculate the energy transfer in coherent neutrino-neutron scattering inside an NS, we model the NS as a homogeneous sphere of radius $R_{\rm NS}\simeq 12.6$ km and mass $M_{\rm NS}\simeq 1.5M_\odot$, where $M_\odot=2\times 10^{33}$g is the solar mass. Since very small momentum transfers are involved, we can neglect the form factor dependence. Since the mean free path of neutrinos is of order cm (see Appendix~\ref{app:D} for detail), most of the scatterings happen in the crust, so the detailed structure of the NS is also not relevant for our study. We also assume a homogeneous energy density distribution with $\rho_n\simeq 7.8\times 10^{14}\:{\rm g}/{\rm cm}^3$. Neutrons form a highly degenerate Fermi sea inside the NS with a Fermi momentum of $p_{f,n}\simeq 450$ MeV. Note that although there is some uncertainty in these numbers depending on the exact equation of state used~\cite{Bell:2019pyc}, for concreteness, we have just used one benchmark equation of state (EoS) for cold non-accreting NS with Brussels–Montreal functionals (BSk24) ~\cite{Pearson:2018tkr}. The microscopic properties of the NS core, species abundances and Fermi momentum, have been averaged over the volume.  

A neutron star moving in the sea of relic neutrinos gravitationally captures them within an impact parameter~\cite{Goldman:1989nd}:
\begin{align}
    \label{eqn: impact parameter}
    b_\mathrm{max} = \left(\frac{2GM_\mathrm{NS}R_\mathrm{NS}}{v_\mathrm{rel}^2}\right)^{1/2}\left( 1 - \frac{2G M_\mathrm{NS}}{R_\mathrm{NS}} \right)^{-1/2}.
\end{align}
where $G$ is Newton's gravitational constant and $v_\mathrm{rel}$ is the relative velocity of the NS traveling in the comoving frame of the Universe. For relic neutrinos with $m_\nu=0.1$ eV, we find $\langle v_\mathrm{rel}\rangle \sim \beta_\nu\simeq \sqrt{3T_\nu/2m_\nu} \simeq 1.9\times 10^4\si{\ km/s}$. Eq.~\eqref{eqn: impact parameter} can be translated into the number of relic neutrinos captured by a NS per unit time as
\begin{align}
    \label{eqn: capture rate}
    \dot{N} &= \pi b_\mathrm{max}^2 v_\mathrm{rel} n_\nu \, ,
\end{align}
where $n_\nu\simeq 56~{\rm cm}^{-3}$ is the total number density of relic neutrinos today, per degree of freedom. Only the neutrinos are captured and participate in cooling of the NS, so we use $3n_\nu$ as the total number density, assuming Dirac neutrinos. The antineutrinos are not included here, because they are reflected near the NS surface. This is due to the fact that the neutron star creates an attractive (repulsive) weak potential for (anti)neutrinos \cite{Arvanitaki:2022oby,Huang:2024tog,Kalia:2024xeq},
\begin{equation}
    U_{\text{weak}} = \frac{G_F}{2\sqrt{2}} n_{\text{matter}} \times
\left\{
\begin{array}{ll}
(-)(3Z - A) & \text{for } \nu_e \ (\bar{\nu}_e) \\
(-)(Z - A) & \text{for } \nu_{\mu,\tau} \ (\bar{\nu}_{\mu,\tau})
\end{array}
\right. \, , 
\end{equation}
where $n_{\text{matter}}$ is the number density of nuclei in the material. For pure neutron matter at the surface of an NS, $ U_{\text{weak}} \approx 13.2~ \text{eV}\left(\frac{\rho}{7 \times 10^{14}{\rm g}/{\rm cm}^3}\right)$ which is much larger than the gravitational attractive potential. Hence the anti-neutrinos will be reflected near the NS surface. As the fraction of anti-neutrinos reflected is controlled by the density gradient near the surface of the NS, our assumption should be considered as a conservative estimate on the cooling of NS.

Eqs.~\eqref{eqn: impact parameter} and \eqref{eqn: capture rate} clearly indicate that only non-relativistic neutrino mass eigenstates are efficiently captured by an NS. 
Gravitational clustering could enhance the local C$\nu$B number density around the NS, but only by a factor of ${\cal O}(1)$~\cite{Ringwald:2004np,Mertsch:2019qjv, Zimmer:2023jbb, Holm:2024zpr}. A gravitationally captured relic neutrino gains energy while rolling down the gravitational potential of the NS. The energy of the relic neutrino on the NS surface is $E_{\nu,\mathrm{surface}} = m_\nu (1 - 2GM_\mathrm{NS}/R_\mathrm{NS})^{-1/2} + K^\infty_\nu$ where $K^\infty_\nu\simeq (3/2)T_\nu$ is the kinetic energy of the relic neutrino at infinite distance from the NS. In the case of light relic neutrinos (with $m_\nu\lesssim 0.1$ eV) where their kinetic energy at the NS surface, $K_\nu=E_{\nu,\mathrm{surface}}-m_\nu\simeq 0.3 m_\nu \ll T_\mathrm{NS}$, energy is transferred from the NS to relic neutrinos in the coherent scattering process, i.e., the NS undergoes anomalous cooling. Numerically, we find that the neutrinos gain about 1\% of their original energy in each scattering (see Appendix~\ref{app:aven}), and  can take away a maximum energy $\simeq 3.15T_\mathrm{NS}$ after a large number of coherent scatterings with the neutrons.  Note that since the neutrons are highly degenerate inside the NS, energy transfer to neutrinos is Pauli-blocked, and the maximum energy transfer will be of order $3T_{\rm NS}$, much smaller than the typical Fermi energy $E_{f,n}$ of the neutrons. Since the neutrino energy on an average increases after each scattering, it is conceivable that a fraction of neutrinos escape after a few interactions and not thermalize. We elaborate this in Appendix~\ref{app:C} and show that, $\sim 25\%$ of the neutrinos go through fifty scatterings, enough to effectively thermalize. The escape timescale after thermalization is much shorter than the age of the NS, as estimated in Appendix~\ref{app:D}. Based on these calculations, we assume that only $25\%$ of the captured neutrinos take away the maximum energy $\sim 3.15 T_{\rm NS}$ and the rest do not effectively participate in the cooling process. 
Thus, the energy loss rate of a NS due to C$\nu$B coherent scattering is given by
\begin{align}
    L_{\rm{C}\nu{\rm B}} &\simeq  \frac{1}{4}\dot{N} \times 3 T_\mathrm{NS} \min\left(1,\frac{\langle \sigma\rangle }{\sigma_{\rm th}}\right), \label{eq:cool}
\end{align}
where $\langle\sigma\rangle$ is the thermally averaged neutrino-NS scattering cross-section that includes the phase-space integration, coherent enhancement factor $N_C$, and Pauli-suppression factor $T_\mathrm{NS}/p_f$; see Appendix~\ref{app:B} for details. The averaged neutrino-NS cross-section is
\begin{align}
\label{Eq: sigma_total}
    \langle\sigma\rangle \simeq 10^{-38}~{\rm cm}^{2}\left(\frac{m_\nu}{0.1 \si{\ eV}}\right)^2\left(\frac{0.1 \si{\ eV}}{|{\bf q}|}\right)^3\left(\frac{T_{\rm NS}}{0.1 \si{\ eV}}\right)
\end{align}
Here $\sigma_{\rm th}$ is the threshold cross-section (see e.g., Ref.~\cite{Baryakhtar:2017dbj})
\begin{align}
    \label{eqn: threshold cross-section}
    \sigma_{\rm th} &\simeq  \frac{\pi R_\mathrm{NS}^2 m_n}{M_\mathrm{NS}}\left(\frac{3T_{\rm NS}-K_\nu}{\left<\Delta E\right>}\right) \nonumber \\
    & \simeq 1.8 \times 10^{-45}\si{\ cm^2} \nonumber \\
    &\ \ \ \times \left(\frac{R_\mathrm{NS}}{10\si{\ km}}\right)^2 \left(\frac{1.5M_\odot}{M_\mathrm{NS}}\right)\left(\frac{3T_{\rm NS}-K_\nu}{\left<\Delta E\right>}\right) \, , 
\end{align}
which corresponds to the cross-section that is enough for neutrinos to obtain kinetic energy $\simeq 3 T_\mathrm{NS}$ after a series of collisions, with the average energy transfer given by $\langle \Delta E\rangle$; see Appendix~\ref{app:aven}.  
Note that the energy loss rate is independent of the neutrino mass, as long as its kinetic energy is much smaller than the NS temperature. We see that $\sigma_{\rm th}\ll \langle\sigma\rangle$, i.e. the coherent neutrino-neutron cross-section is much larger than the threshold value, which satures the limit in Eq.~\eqref{eq:cool}.  Therefore, adding the neutrino-electron scattering will not change our result for the cooling effect. 

To see whether the anomalous cooling rate given by Eq.~\eqref{eq:cool} is observable or not, we have to compare it with the standard photon cooling rate (for a fixed NS radius): 
\begin{align}
    L_\gamma = 4\pi R_{\rm NS}^2\sigma_{\rm SB} T_{\rm NS}^4 \, ,
\end{align}
where $\sigma_{\rm SB}$ is the Stefan-Boltzmann constant. We thus find that the ratio of the kinetic cooling rate from C$\nu$B to the black-body cooling rate from photon emission is

\begin{align}
\label{eq: anomalous_cooling_rate}
    \frac{L_{\rm{C}\nu{\rm B}}}{L_\gamma} &= \frac{3 G M_\mathrm{NS}n_\nu}{4 \sigma_{\rm SB} R_\mathrm{NS} T_\mathrm{NS}^3 v_\mathrm{rel}} \left(1 - \frac{2G M_\mathrm{NS}}{R_\mathrm{NS}}\right)^{-1}\nonumber \\
    & \simeq \sum_i 6 \times 10^{-10} \left(\frac{10^3 \si{\ K}}{T_\mathrm{NS}}\right)^3 \left(\frac{0.1 \si{\ eV}}{m_i}\right)^{1/2}.
\end{align}
where the sum includes all the non-relativistic neutrino mass eigenstates.

Thus, we find that the energy loss induced by C$\nu$B scattering is subdominant compared to that due to photon emission. The anomalous cooling can ostensibly be observable if there exists a local C$\nu$B overdensity $\gtrsim 10^9$. Such large overdensity locally is still allowed by laboratory constraints like KATRIN~\cite{KATRIN:2022kkv},  but has to be reconciled with other observational constraints and projections~\cite{Brdar:2022kpu, Bauer:2022lri, Tsai:2022jnv, Ciscar-Monsalvatje:2024tvm,Franklin:2024enc,DeMarchi:2024zer,Herrera:2024upj}. Large overdensities are  theoretically possible to achieve, e.g. by forming neutrino-antineutrino bound states via an attractive Yukawa potential~\cite{Smirnov:2022sfo}. However, the neutrino bound states in the overdense cluster behaves as a degenerate Fermi gas that has very diferent capture rate by an NS compared to Eq.~\eqref{eqn: impact parameter} and produces different phenomenology than continuous anomalous cooling as in Eq.~\eqref{eq: anomalous_cooling_rate}. 
\section{NS Heating by Sterile Neutrinos}
\label{sec: NS_heating}
For coherent scattering of heavier particle species with the neutrons in the NS, the situation is very different. In particular, if the energy of the incoming particle is larger than $T_{\rm NS}$, it will deposit almost all of its energy in the NS and get captured, thus heating the NS. However, given the stringent upper bound on the active neutrino mass $m_\nu\lesssim 0.1$ eV (which corresponds to $\sim 1000$ K) from cosmological~\cite{Planck:2018vyg, DESI:2024mwx} as well as laboratory data~\cite{Katrin:2024tvg}, it is unlikely for C$\nu$B to heat up the NS to an observable extent. We should clarify here that even if the cosmological upper limit on the sum of neutrino masses could in principle be evaded in the present Universe by invoking some non-standard cosmology or time-varying neutrino masses~\cite{Fardon:2003eh, Krnjaic:2017zlz, Lorenz:2018fzb,Huang:2022wmz,Goertz:2024gzw}, the laboratory bounds on $\left(m_{\nu_\alpha}^{\rm eff}\right)^2  \equiv \sum_i |U_{\alpha i}|^2m_i^2$ (where $U$ is the PMNS mixing matrix and $m_i$ are the three active neutrino mass eigenvalues) are unavoidable. The most stringent bound is on the electron neutrino mass, $ m_{\nu_e}^{\rm eff}< 0.45~{\rm eV}$ from KATRIN~\cite{Katrin:2024tvg}, while the corresponding bounds on muon and tau neutrino masses are much weaker: $m_{\nu_\mu}^{\rm eff}< 0.19~{\rm MeV}$ and $m_{\nu_\tau}^{\rm eff}< 18.2~{\rm MeV}$~\cite{ParticleDataGroup:2022pth}. 
However, after taking into account the observed neutrino oscillation data~\cite{nufit}, there is no way one could make any of the mass eigenvalues $m_i$ larger than the KATRIN upper limit on $m_{\nu_e}^{\rm eff}$, thus rendering the weaker upper limits on $m_{\nu_{\mu,\tau}}^{\rm eff}$ practically irrelevant.   

\begin{figure}[t!]
    \centering
    \includegraphics[width=0.45\textwidth]{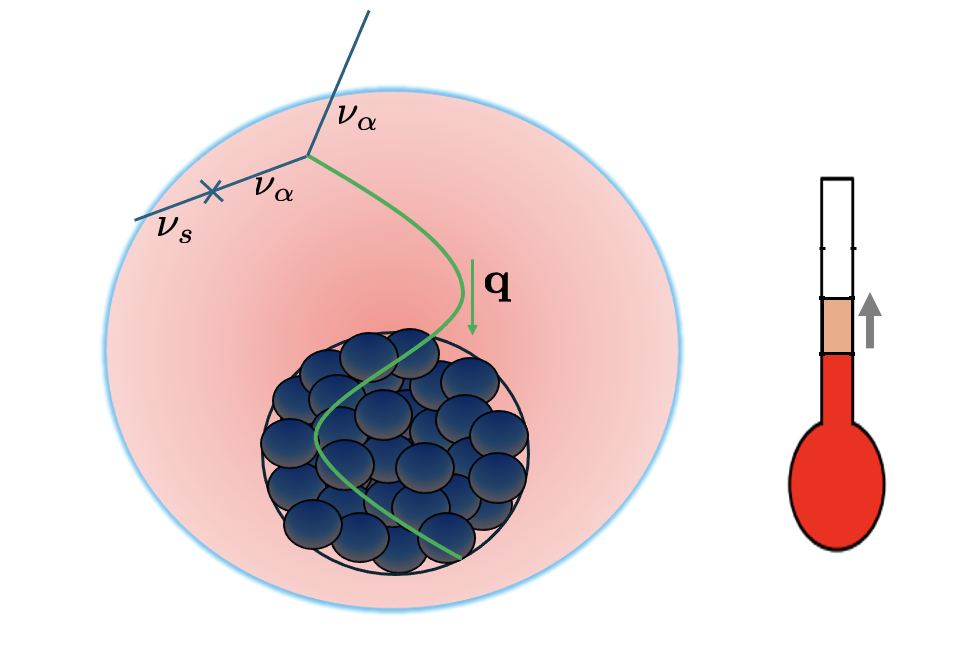}
    \caption{An artist's rendition of the NS kinetic {\it heating} via coherent scattering of sterile neutrino DM $\nu_s$, which converts to an active neutrino $\nu_\alpha$. }
    \label{fig:artist2}
\end{figure}

On the other hand, the anomalous heating of NS could be induced by some heavier beyond the Standard Model (BSM) particles, such as DM. The dark kinetic heating of NS has been extensively studied before for $\gtrsim$ GeV-scale DM~\cite{Kouvaris:2007ay, Kouvaris:2010vv, deLavallaz:2010wp, Bramante:2017xlb, Raj:2017wrv, Baryakhtar:2017dbj, Bell:2018pkk, Joglekar:2019vzy, Joglekar:2020liw, Bell:2019pyc, Bell:2020jou, Bell:2020lmm, Chatterjee:2022dhp, Avila:2023rzj,Bell:2023ysh}; see Ref.~\cite{Bramante:2023djs} for a review.  Here we examine for the first time the NS kinetic heating effect for a keV-scale sterile neutrino DM; see Fig.~\ref{fig:artist2}. Sterile neutrinos are well-motivated BSM particles, originally introduced to explain the non-zero neutrino mass by the seesaw mechanism~\cite{Minkowski:1977sc, Mohapatra:1979ia, Yanagida:1979as, Gell-Mann:1979vob}. In the minimal scenario, the sterile neutrinos can only interact with the SM sector via their mixing with the active neutrinos. Depending on the smallness of this mixing angle $\theta$, sterile neutrinos can be cosmologically stable and can serve as a warm/cold DM~\cite{Asaka:2005an, Drewes:2016upu}. We assume the sterile neutrinos to make 100\% of the DM, with their energy density given by the local DM density in the galactic halo, $\rho_{\nu_s}\simeq 0.4\si{\ GeV/cm^3}$~\cite{Callingham:2018vcf}. The DM density can be much larger closer to the galactic center; however, since the kinetic heating effect we are interested in can be detectable only for nearby NSs (within a few pc from Earth), the larger DM density at the galactic center does not help our signal.    

The sterile neutrinos can undergo coherent scattering off neutrons in the NS in the same way as the active neutrinos~\cite{Ando:2010ye}, but now with cross-sections suppressed by  $\sin^2\theta$, where $\theta$ is the mixing angle between the sterile neutrino and the electron neutrino.\footnote{For mixing with muon or tau neutrino, the charged-current scattering with electrons is not applicable, and only the neutral-current scattering with the nucleons is relevant.} For keV-scale sterile neutrinos, their kinetic energy is larger than $T_{\rm NS}$, and therefore, in each collision, they will lose a significant fraction of their energy to the NS, and the remaining energy will be transferred to the converted active neutrino. We checked that the mean free path of the produced active neutrino is short enough to transfer almost all of the kinetic energy to the NS, provided the sterile neutrino mass is below MeV-scale. In this process, the energy transferred by a sterile neutrino to the neutron star is the sum of kinetic energy and mass energy (assuming the NS temperature to be much smaller than the sterile neutrino mass). This is reminiscent of the NS heating by DM annihilation~\cite{Kouvaris:2007ay} which can raise the observed NS surface temperature to larger values compared to the case when only the kinetic energy is transferred. Another similarity with the DM annihilation should also be noted here:  Since the coherence enhancement factor in Eq.~\eqref{eq:3}, hence the scattering cross-section between the active neutrino and neutron, increases rapidly with decreasing sterile neutrino mass, the energy transfer does not suffer from Pauli blocking unlike the dark kinetic heating~\cite{Baryakhtar:2017dbj}, for which the energy transfer can become inefficient for DM masses below the neutron Fermi momentum. We can then calculate the total energy gain rate of the NS just as a product of the number of captured sterile neutrinos in unit time and the energy transferred to a NS by each sterile neutrino. Discussion of energy gain without this approximation can be found, e.g., in Ref.~\cite{Bell:2018pkk}. Note that the mean free path of sterile neutrinos inside the NS will be increased by a factor of $\sin^{-2}\theta$, compared to the relic neutrinos with the same energy. When the mean free path becomes larger than the size of the star by a given factor (say $x$), only a fraction ($x$) of the sterile neutrinos will interact with the star. This has been taken into account in our calculation.

We find the energy gain of a neutron star per unit time is as follows:
\begin{align}
    L_{\nu_s} &= \dot{N} (E_{\nu_s,\mathrm{surface}} - 3 T_\mathrm{NS}) \min\left(1,\frac{\langle\sigma\rangle}{\sigma_{\rm th}}\right) ,
\end{align}
where $\dot{N}$ is given by Eq.~\eqref{eqn: impact parameter} with $n_\nu$ replaced by $n_{\nu_s}=\rho_{\nu_s}/m_{\nu_s}$, $v_\mathrm{rel} = 330.5\si{\ km/s}$ which is the average relative velocity for a Maxwellian velocity distribution of DM with mean value $v_{\rm DM}=230~{\rm km/s}$ and $v_{\rm NS}=207~{\rm km/s}$ by following Ref.~\cite{2009PASP..121..814O} which studied the velocity distribution of old NSs by performing numerical simulations. $\langle \sigma\rangle$ is the thermally averaged scattering cross-section given in Appendix~\ref{app:B} scaled by $\sin^2\theta$, and $E_{\nu_s,{\rm surface}}=m_{\nu_s}\left(1 - 2G M_\mathrm{NS}/R_\mathrm{NS}\right)^{-1/2}$ is the energy of the sterile neutrino on the NS surface, assuming that its kinetic energy at infinity is negligible. The threshold cross-section in this case will be slightly modified from Eq.~\eqref{eqn: threshold cross-section}: 
\begin{align}
    \label{eqn: threshold cross-section2}
    \sigma_{\rm th} &= \frac{\pi R_\mathrm{NS}^2 m_n}{M_\mathrm{NS}}\max\left(\frac{E_{\nu_s,\mathrm{surface}} - 3 T_\mathrm{NS}}{\left<\Delta E\right>},1\right) \, .
 \end{align}
The active neutrinos being produced from the sterile neutrino interactions (see Fig.~\ref{fig:artist2}) will deposit most of their kinetic energy to the NS, but as long as  $m_{\nu_i}$ is smaller than $0.1 T_{\rm NS}$, they can still escape the gravitational potential of the NS and do not form a Fermi sphere inside.  Here we include the  electron scattering for the NC channel, which contributes about 20\%. The CC electron scattering effect is much smaller, because of the large momentum transfer, which suppresses the coherent enhancement.

\begin{figure*}[t!]
    \centering
    \includegraphics[width=0.9\linewidth]{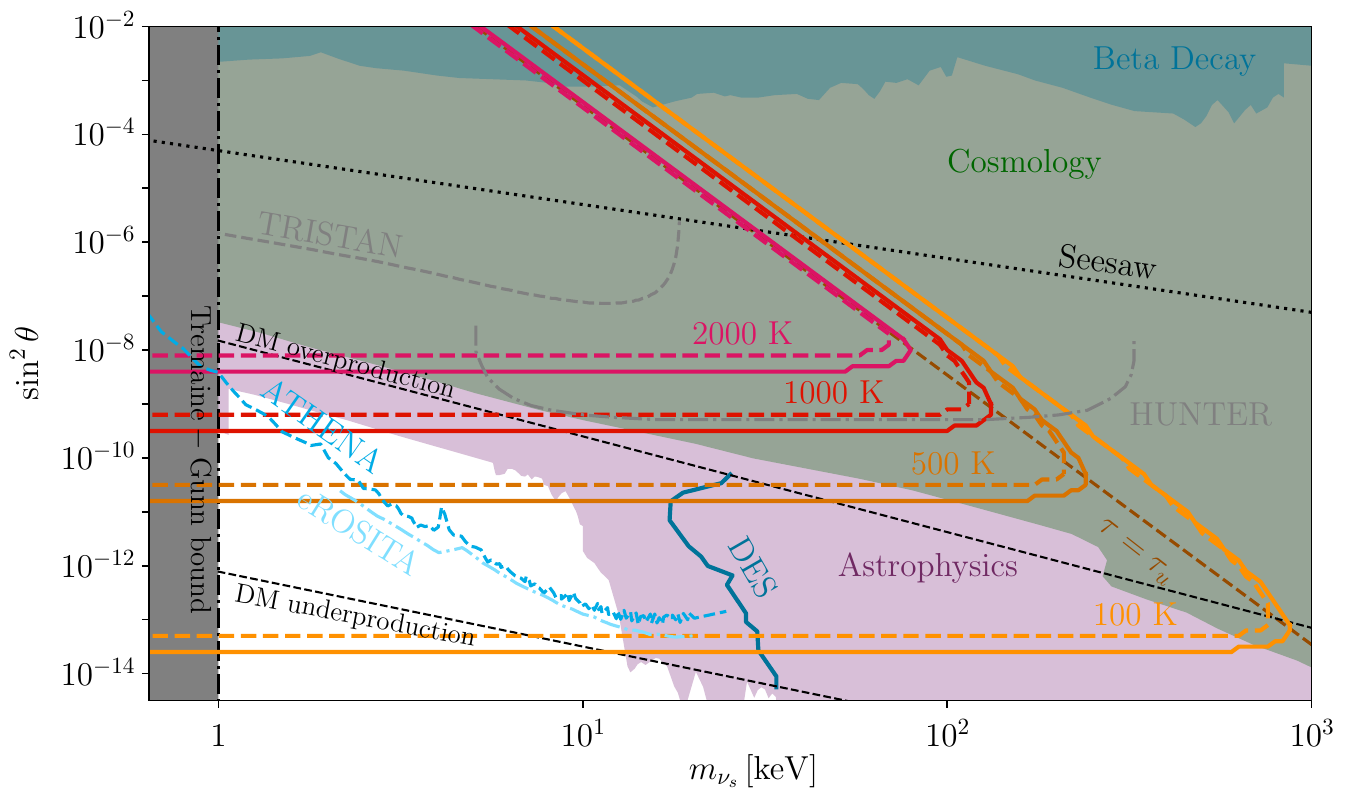}
    \caption{\label{fig: constraints for mixing with sterile neutrinos}\raggedright Projected limits on sterile neutrino mass and mixing from NS kinetic heating observations at different temperatures. The solid and dashed curves represent the variation due to the EoS parameters. The shaded regions are excluded, while the dashed lines show future sensitivities of complementary probes. See text for details.}  
    \label{fig:sterile}
\end{figure*}

Including the effect of gravitational redshift, the effective NS surface temperature induced by the kinetic heating is given by
\begin{align}
    4\pi R_\mathrm{NS}^2 \sigma_{\rm SB} T_\mathrm{heating}^4 &= L_{\nu_s} \left(1 - \frac{2GM_\mathrm{NS}}{R_\mathrm{NS}}\right).
\end{align}
This gives the modified surface temperature observed at infinity: 
$T_\infty = (T^4_\mathrm{NS} + T^4_\mathrm{heating})^{1/4}$. In Fig.~\ref{fig:sterile}, we show different contours of  $T_\infty$ in the sterile neutrino mass-mixing plane. The solid and dashed contours correspond to two different values of the central mass-energy density that determines the mass-radius relation for the same EoS, namely, BSk24-1 and BSk24-2 parameters from Refs.~\cite{Pearson:2018tkr, Bell:2019pyc}.\footnote{The other EoS (BSk25) used in Refs.~\cite{Pearson:2018tkr, Bell:2019pyc} gives sensitivities very similar to the ones shown in Fig.~\ref{fig:sterile}.} The benchmark of 2000 K is motivated by the fact that the corresponding black-body spectrum from an NS located 10 pc from Earth yields a spectral flux density of $f_\lambda\simeq 0.5$ nJy at wavelength $\lambda\sim 2~\mu$m, which should be detectable by JWST~\cite{Gardner:2006ky}. The 1000 K benchmark is at the very edge of the detection threshold for JWST, but within reach of future telescopes like ELT and TMT. The 500 K benchmark may lie at the detection threshold for future telescopes, while the 100 K benchmark is very difficult to achieve and is shown here for illustration purpose only.    

The dominant decay channel of sterile neutrinos with mass smaller than twice of electron mass is $\nu_s \to \nu_\alpha \nu_\beta \bar{\nu}_\beta$ whose decay width is given by~\cite{Pal:1981rm, Barger:1995ty}
\begin{align}
    \Gamma_{\nu_s \to \nu_\alpha\nu_\beta \bar{\nu}_\beta} & = \frac{G_F^2 m_{\nu_s}^5}{96\pi^3} \sin^2\theta \nonumber \\
    & \simeq  \frac{1}{\tau_u} \left(\frac{m_{\nu_s}}{50\si{\ keV}}\right)^5 \left(\frac{\sin^2\theta}{1.1 \times 10^{-7}}\right) ,
\end{align}
where $\tau_u = 13.8\si{\ Gyr}$~\cite{Planck:2018vyg} is the age of the Universe. This is shown in Fig.~\ref{fig:sterile} by the diagonal brown dashed line labeled `$\tau=\tau_u$'. The NS heating curves run close to the $\tau=\tau_u$ line on the upper part of Fig.~\ref{fig:sterile}, because to the right of this line, the sterile neutrino decays deplete its abundance rapidly and there are not enough of them to interact with the NS. In the lower part of the Figure, the sensitivities are independent of mass, because we have fixed the DM energy density $\rho_{\nu_s}$, which implies that the number density $n_{\nu_s}$ goes inversely as $m_{\nu_s}$. The total energy transfer is proportional to $n_{\nu_s}m_{\nu_s}$ which is therefore independent of $m_{\nu_s}$. 

There is another radiative decay mode $\nu_s\to \nu_\alpha\gamma$, which is suppressed by a factor of $27\alpha/8\pi$ with respect to the $3\nu$ decay mode~\cite{Pal:1981rm}; however, since this is a 2-body decay with a single photon in the final state having energy at half the sterile neutrino mass, monochromatic $X$-ray~\cite{Ng:2019gch, Roach:2019ctw}/$\gamma$-ray~\cite{Laha:2020ivk} line searches put a stringent constraint using this channel, as shown by the purple-shaded region labeled `Astrophysics'. Future $X$-ray missions like ATHENA~\cite{Neronov:2015kca} and eROSITA~\cite{Dekker:2021bos} will further improve these bounds, as shown by the dashed blue curves. The `Astrophysics' constraint at lower masses also includes the supernova cooling~\cite{Shi:1993ee}. Similarly, cosmological observations including $\Delta N_{\rm eff}$, BAO and $H_0$ exclude the green-shaded region labeled `Cosmology'~\cite{Vincent:2014rja,Bridle:2016isd}. There also exist constraints from Lyman-$\alpha$~\cite{Garzilli:2019qki, Irsic:2023equ} and Milky Way subhalos~\cite{Horiuchi:2013noa, Dekker:2021scf} and satellite galaxies~\cite{DES:2020fxi}, all of which rule out sterile neutrino DM below a few keV. We show here the latest DES+PS1 bound (labeled `DES')~\cite{DES:2020fxi}, but although stronger than the theoretical Tremaine-Gunn bound obtained from phase-space considerations~\cite{Tremaine:1979we}, it depends on the sterile neutrino DM production mechanism~\cite{Dodelson:1993je, Shi:1998km}, and can in principle be modified in presence of additional interactions~\cite{An:2023mkf, Astros:2023xhe}; therefore, we do not shade this region. For the same reason, we do not shade the DM over/underproduction regions, taken from Ref.~\cite{Schneider:2016uqi}, which could get modified, e.g. in presence of active~\cite{An:2023mkf} or sterile self-interactions~\cite{Astros:2023xhe}. 
Finally, for sterile neutrino mixing with electron flavor, there exist constraints from beta-decay spectrum in different nuclei~\cite{Atre:2009rg, Bolton:2019pcu}, as shown by the cyan-shaded region labeled `Beta Decay'. Future beta-decay experiments like TRISTAN~\cite{KATRIN:2018oow} and HUNTER~\cite{Martoff:2021vxp} can improve these bounds significantly, as shown by the corresponding dashed curves. The `Seesaw' line shown here is the suggestive value of the mixing angle for a given sterile neutrino mass to get the correct neutrino mass via type-I seesaw mechanism~\cite{Minkowski:1977sc, Mohapatra:1979ia, Yanagida:1979as, Gell-Mann:1979vob}.  From Fig.~\ref{fig:sterile}, it is clear that NS kinetic heating observations could put new meaningful constraints on the sterile neutrino DM scenario, which would be complementary to the existing astrophysical, cosmological, and laboratory probes. 

Note that in Fig.~\ref{fig:sterile}, we show the constraints as a function of $\sin^2\theta$, not $\sin^2(2\theta)$, as sometimes done in the literature. This difference arises because the mixing probability between a mass eigenstate and a flavor eigenstate is given by $\sin^2\theta$, whereas the mixing probability between two flavor states is given by $\sin^2\theta\cos^2\theta=\sin^2(2\theta)/4$ (because of the mixing matrix being inserted twice). Since the sterile neutrino DM is considered here to be in its mass eigenstate, it is more appropriate for us to use $\sin^2\theta$ as the mixing probability. We have translated the other constraints accordingly from $\sin^2(2\theta)$ to $\sin^2\theta$. 

Also note that Fig.~\ref{fig:sterile} is given for the sterile neutrino mixing with the electron flavor. For mixing with muon or tau flavor, there will be a few differences, e.g. in the supernova bounds~\cite{Arguelles:2016uwb}, and no $\beta$-decay constraints. Moreover, there will be no neutrino-electron charged-current scattering contribution to the NS heating and only the neutrino-nucleon neutral-current scattering should be considered. However, since we have neglected the electron contribution in our calculation (see Appendix~\ref{app:A}), it practically makes no difference to our sensitivity results shown in Fig.~\ref{fig:sterile}.

\section{Summary}
\label{sec: summary}
Detection of relic neutrinos is an extremely important challenge for Particle Physics. Recognizing that  these are very low energy neutrinos  and
therefore their associated de Brogile wavelength is of macroscopic size,
we have tried to exploit the enormous density of neutrons in a neutron star via
coherent scattering. Unfortunately, still, cooling of NS via relic neutrino emission
is subdominant to thermal radiation. However, sterile neutrinos through their mixing with active neutrinos
cause anomalous heating which may well have observational possibilities.


The future detectability of the kinetic cooling (heating) effect studied here relies on the crucial assumption that there is no other active cooling/heating mechanism in play so that we can identify the anomalous cooling (heating) as solely due to the coherent scattering of relic neutrinos (sterile neutrino DM). In practice, NSs cooling passively since their birth may as well be affected by some other late-time reheating mechanism, either of astrophysical origin~\cite{2010A&A...522A..16G}, such as magnetic field decay, rotochemical heating, vortex creep heating~\cite{1984ApJ...276..325A,1984ApJ...278..791A,Fujiwara:2023tmr,Fujiwara:2023hlj},  and crust-cracking, or of BSM physics origin~\cite{Bramante:2023djs}, such as accretion of heavier DM and baryon-number violating (di)nucleon decays. These additional heating mechanisms could complicate interpretation; however, dedicated future observational campaigns to measure thermal luminosities of old NSs might resolve some of these ambiguities~\cite{Raj:2024kjq}.

In addition, our results could be affected by theoretical uncertainty in the composition of and dynamics in the NS core. 
Many theoretical models expect the emergence of exotic phases of matter such as hyperon~\cite{1960SvA.....4..187A} and color superconducting phases~\cite{Alford:2002rj}. Strong interaction in the matter with a density above the saturation density is quite difficult to understand. Moreover, an improved calculation of the coherence effect using dynamic structure functions is desirable, since we are dealing with an interacting quantum many-body system~\cite{PhysRev.95.249}. We leave these directions for future work.  
\section*{Acknowledgments}
We thank Mark Alford, Liam Brodie, Chris Cappiello, Garv Chauhan, Vijai Dixit, Alexander Haber, Steven Harris, Shmuel Nussinov and Flip Tanedo for useful discussions and/or comments on the draft. We especially thank Garv Chauhan for informing us about his concurrent work on a related topic~\cite{Chauhan:2024, Chauhan:2024deu}. S.D. is partly supported by the McDonnell Center for the Space Sciences. B.D. and T.O. are partly supported by the U.S. Department of Energy under grant No.~DE-SC 0017987.  T.O. would like to thank the University Research Association Visiting Scholars Program. B.D. and T.O. thank the Fermilab Theory Group for warm hospitality where this work was completed. B.D. acknowledges the Center for Theoretical Underground Physics and Related Areas (CETUP* 2024) and the Institute for Underground Science at SURF for hospitality and for providing a stimulating environment, where a part of this work was done.

\appendix
\onecolumngrid
\section{Neutrino-nucleon and neutrino-electron cross sections}
\label{app:A}
\begin{figure}[h!]
    \centering
    \includegraphics[width=0.8\linewidth]{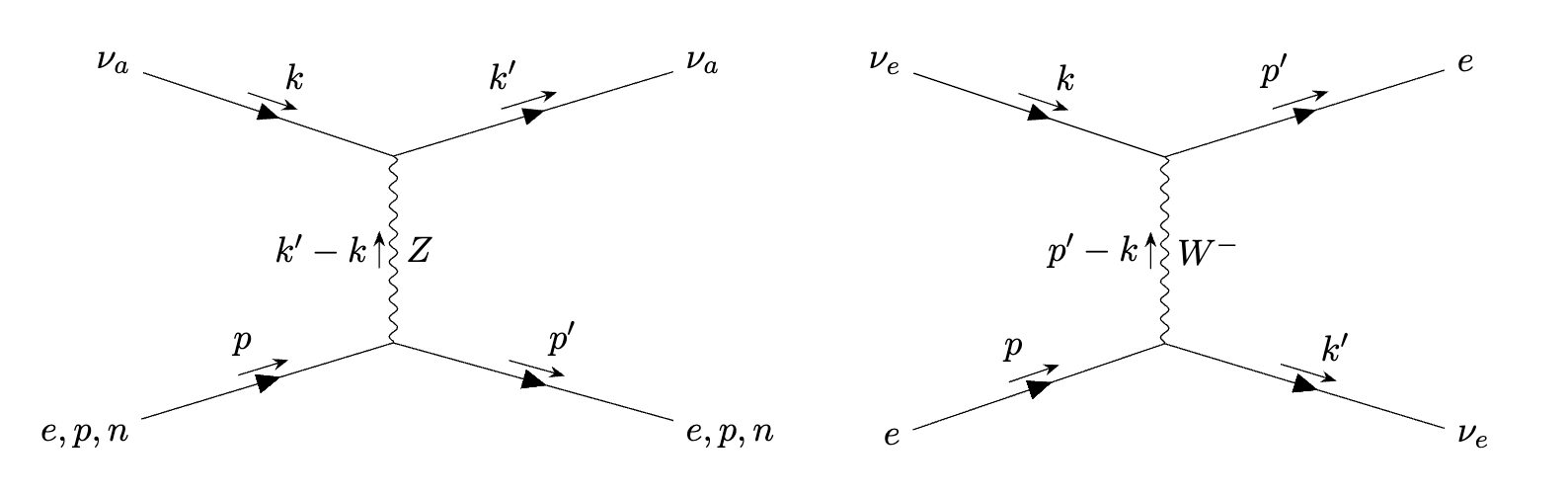}
    \caption{Feynman diagrams of relevant neutral-current (left) and charged-current (right) neutrino scattering processes.}
    \label{fig:feyn}
\end{figure}

In this section, we give the details of the neutrino-nucleon and neutrino-electron scattering cross section calculations. The corresponding Feynman diagrams are shown in Fig.~\ref{fig:feyn}. 

The elastic neutrino-nucleon scattering proceeds via neutral current $Z$ exchange which does not change the quark flavor of the nucleon, and is the same for all neutrino flavors. Assuming the nucleons are non-relativistic, the differential $\nu-N$ cross section is given by~\cite{Formaggio:2012cpf, Leitner:2005} 
\begin{align}
    \frac{d\sigma_{\nu N}}{dQ^2} = \frac{m_N^2G_F^2}{8\pi E_\nu^2}\left[A-\frac{(s-u)}{m_N^2}B+\frac{(s-u)^2}{m_N^4}C\right] \, , 
\end{align}
with $m_N$ being the nucleon mass (with $N=p,n$), $Q^2=-q^2$ is the momentum transfer which equals the Mandelstam variable $t$, $s-u=4m_NE_\nu-Q^2$, and $E_\nu$ is the incoming neutrino energy. The factor $A,B,C$ are functions of $\tau=Q^2/4m_N^2$ and the form factors: 
\begin{align}
    A&=4\tau\left[(1+\tau)\left(\tilde{F}_A^N\right)^2-(1-\tau)\left(\tilde{F}_1^N\right)^2+\tau(1-\tau)\left(\tilde{F}_2^N\right)^2+4\tau\tilde{F}_1^N\tilde{F}_2^N\right] \, , \\
    B&=4\tau\tilde{F}_A^N\left(\tilde{F}_1^N+\tilde{F}_2^N\right) \, , \\
    C&=\frac{1}{4}\left[\left(\tilde{F}_A^N\right)^2+\left(\tilde{F}_1^N\right)^2+\tau\left(\tilde{F}_2^N\right)^2 \right] \, .
\end{align}
The proton and neutron form factors are given as follows:
\begin{align}
    \tilde{F}_{1,2}^p &= \frac{1}{2}\left[(1-4\sin^2\theta_w)F_{1,2}^p-F_{1,2}^n-F_{1,2}^S\right] \, , \\
     \tilde{F}_{1,2}^n &= \frac{1}{2}\left[(1-4\sin^2\theta_w)F_{1,2}^n-F_{1,2}^p-F_{1,2}^S\right] \, , 
\end{align}
with the Dirac and Pauli form factors $F_1^{p,n}$ and $F_2^{p,n}$ related to the Sachs form factors:
\begin{align}
    F_1^{p,n} & =\frac{G_M^{p,n}+\tau G_E^{p,n}}{1+\tau} \, , \\
    F_2^{p,n} & = \frac{G_M^{p,n}-G_E^{p,n}}{1-\tau} \, ,
\end{align}
where $G_M^{p,n}$ and $G_E^{p,n}$ are the magnetic and electric form factors of the nucleon, respectively. We will use the BBA-2003 form factor fit~\cite{Budd:2003wb} which gives
\begin{align}
    G_{E,M}^N(Q^2) = \frac{G_{E,M}^N(0)}{1+a_2Q^2+a_4Q^4+a_6Q^6+a_8Q^8+a_{10}Q^{10}+a_{12}Q^{12}} \, ,
    \label{eq:BBA}
\end{align}
where $a_2,\cdots,a_{12}$ are fit parameters listed in Table~\ref{tab:fit} and $G_{E,M}^N(0)$ are the dipole form factors at $Q^2=0$: 
\begin{align}
    G_E^p(Q^2) = G_D(Q^2), \quad G_E^n(Q^2)=0, \quad G_M^p(Q^2)=\mu_p G_D(Q^2), \quad G_M^n(Q^2) = \mu_nG_D(Q^2) \, ,
\end{align}
where $\mu_p=2.793$ and $\mu_n=-1.913$ are the magnetic moments of proton and neutron respectively, and  
\begin{align}
G_D(Q^2) = \frac{1}{\left( 1+\frac{Q^2}{m_V^2}\right)^2} \, ,
\end{align}
with the vector mass $m_V=0.843$ GeV. For the electric form factor of the neutron (which is vanishing in the dipole formalism), we use the following parametrization~\cite{Krutov:2002tp}:
\begin{align}
    G_E^n(Q^2) = -\mu_n\frac{a\tau}{1+b\tau}G_D(Q^2) \, ,
\end{align}
with $a=0.942$ and $b=4.61$.
\begin{table}[h!]
\begin{tabular}{|c|c|c|c|c|c|c|}\hline
     &  $a_2$ &  $a_4$&  $a_6$&  $a_8$&  $a_{10}$&  $a_{12}$\\ \hline
    $G_E^p$ & 3.253 & 1.422 & 0.08582 & 0.3318 & $-0.09371$ & 0.01076 \\
    $G_p^M$ & 3.104 & 1.428 & 0.1112 & $-0.006981$ & 0.0003705 & $-0.7063\times 10^{-5}$ \\
    $G_M^n$ & 3.043 & 0.8548 & 0.6806 & $-0.1287$ & 0.008912 & 0 \\ \hline
\end{tabular}
\caption{Coefficients $a_i$ (in units of GeV$^{-i}$) of the inverse ponynomial fits for the form factors in Eq.~\eqref{eq:BBA}. }
\label{tab:fit}
\end{table}
The axial form factor is given by 
\begin{align}
    \tilde{F}_A^{p,n}=\frac{1}{2}\left(\pm F_A+F_A^S\right) \, , 
\end{align}
where the charged current axial form factor has $+(-)$ sign for proton (neutron). We use a dipole form
\begin{align}
    F_A(Q^2) = \frac{g_A}{\left(1+\frac{Q^2}{m_A^2} \right)^2} \, ,
\end{align}
with the axial vector constant $g_A=-1.267$. For the strange axial form factors, we use~\cite{Garvey:1992cg} 
\begin{align}
F_1^S(Q^2) & = \frac{F_1^S(0)Q^2}{(1+\tau)\left(1+\frac{Q^2}{m_V^2} \right)^2} \, , \\
F_2^S(Q^2) & = \frac{F_2^S(0)Q^2}{(1+\tau)\left(1+\frac{Q^2}{m_V^2} \right)^2} \, , \\
    F_A^S(Q^2) & = \frac{\Delta s}{\left(1+\frac{Q^2}{m_A^2} \right)^2} \, ,
\end{align}
with $\Delta s=-0.21$, $F_1^S(0)=0.53$, $F_2^S(0)=-0.40$ and $m_A=1.012$ GeV. 

\begin{figure}[t!]
\includegraphics[width=0.7\textwidth]{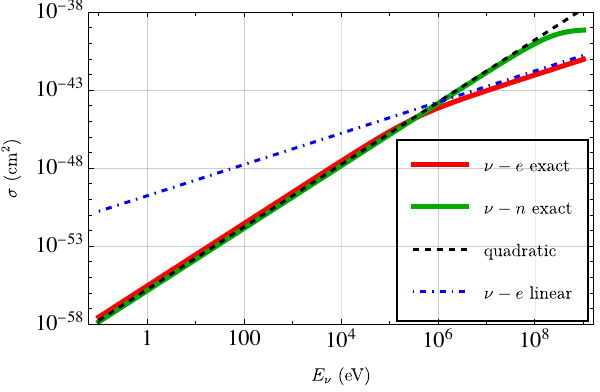}
\caption{Comparison of neutrino-nucleon and neutrino-electron cross sections as functions of the neutrino energy in the rest frame of the target. The linear and quadratic approximations are given in Eqs.~\eqref{eq:lin} and \eqref{eq:quad}, respectively.}
\label{fig:cross}
\end{figure}

The total neutrino-nucleon cross section is given by 
\begin{align}
    \sigma_{\nu N}(E_\nu) = \int_{Q^2_{\rm min}}^{Q^2_{\rm max}}dQ^2 \frac{d\sigma_{\nu N}}{dQ^2} \ , , 
\end{align}
with integration limits 
\begin{align}
    Q^2_{\rm min,max} = \frac{2E_\nu^2m_N\mp E_\nu(s-m_N^2)}{2E_\nu+m_N} \, ,
\end{align}
with $s=m_N^2+2m_NE_\nu$. The result is shown in Fig.~\ref{fig:cross} by the green solid curve. 

For the neutrino-electron scattering, there are both neutral current and charged current contributions. The leading-order differential cross section can be expressed in terms of the final-state neutrino energy $E_\nu'$ as~\cite{Tomalak:2019ibg} 
\begin{align}
    \frac{d\sigma_{\nu e}}{dE'_\nu} = \frac{m_e}{4\pi}\left[c_L^2I_L+c_R^2I_R+c_Lc_RI_R^L\right] \, ,
\end{align}
where $c_L=2\sqrt 2 G_F\left(\sin^2\theta_w+\frac{1}{2}\right)$, $c_R=2\sqrt 2 G_F \sin^2\theta_w$, and the kinematical factors in the elastic limit are 
\begin{align}
    I_L=1 \, , \quad I_R=\left(\frac{E'_\nu}{E_\nu}\right)^2 \, , \quad I_R^L=-\frac{m_e}{E_\nu}\left(1-\frac{E'_\nu}{E_\nu}\right) \,.
\end{align}
The total cross section is given by 
\begin{align}
    \sigma_{\nu e}(E_\nu) = \int_{E'_{\nu, \rm min}}^{E'_{\nu, \rm max}} dE'_\nu \frac{d\sigma_{\nu e}}{dE'_\nu} \, ,
\end{align}
with the limits of integration
\begin{align}
    E'_{\nu, \rm min} = \frac{m_eE_\nu}{m_e+2E_\nu} \, , \quad E'_{\nu, \rm max}=E_\nu \, .
\end{align}
The result is shown in Fig.~\ref{fig:cross} by the red solid curve. Note that for $E_\nu\gg m_e$, the cross section can be well-approximated by a linear term:
\begin{align}
    \sigma_{\nu e}(E_\nu \gg m_e)\simeq \frac{G_F^2}{\pi}2m_eE_\nu \, ,
    \label{eq:lin}
\end{align}
as shown by the dot-dashed line in Fig.~\ref{fig:cross}. On the other hand, for $E_\nu\ll m_e$ (or $E_\nu\ll m_N$ for nucleon case), the cross section can be well-approximated by the quadratic form:
 \begin{align}
    \sigma(E_\nu\ll m_{\rm target})\simeq \frac{G_F^2}{\pi}E_\nu^2 \, , 
    \label{eq:quad}
\end{align}
as shown by the dashed line in Fig.~\ref{fig:cross}. The same is true for antineutrinos. This justifies our use of Eq.~\eqref{eq:crosssim} in the main text. 

From Fig.~\ref{fig:cross}, we find that for the entire energy range of our interest, i.e. sub-eV for the relic neutrino case and few keV for the sterile neutrino case, the $\nu-e$ and $\nu-N$ cross sections are comparable. However, the electron number density in the NS is roughly an order of magnitude smaller than the neutron number density, which makes the electron contribution subdominant in the NS environment. Moreover, most of the electrons are believed to be near the surface of the NS, whereas for the sterile neutrino case with small mixing (hence large mean free path), the scattering is most likely to happen deep inside the NS, thus making the electron scattering even more unlikely.

\section{Coherent scattering of neutrinos inside a neutron star}
\label{app:B}
In this section, we give a detailed analytic evaluation of the neutrino coherent cross-section inside the NS via neutral-current interactions. The mass basis of neutrinos is used throughout this section and we assume Dirac neutrinos. 

In general, for a neutrino-nucleus coherent scattering, assuming that the coherence length is larger than the size of the nucleus, the  squared amplitude of a coherent scattering between a Dirac (anti)neutrino (with energy $E_\nu$ much lower than the rest mass $m_N$) and a single nucleus (with the mass number $A$ and the atomic number $Z$), averaging over target nucleus spins, and summing over the final state helicities, is given by~\cite{Formaggio:2012cpf,Domcke:2017aqj, Bauer:2022lri}
\begin{align}
\begin{aligned}
    \left\langle\left|\mathcal{M}_N\left(\nu_{i, {\sf s}}^D\right)\right|^2\right\rangle & 
    =\frac{G_F^2}{16} \mathcal{T}_{\alpha \beta} \mathcal{T}_X^{\alpha \beta}, \\
    \left\langle\left|\mathcal{M}_N\left(\bar{\nu}_{i, {\sf  s}}^D\right)\right|^2\right\rangle & =\frac{G_F^2}{16} \overline{\mathcal{T}}_{\alpha \beta} \mathcal{T}_X^{\alpha \beta}, \\
\end{aligned}
\end{align}
where {\sf s} is the helicity of the incoming (anti)neutrino mass eigenstate $i$, and the traces  $\mathcal{T}_{\alpha \beta}, \overline{\mathcal{T}}_{\alpha \beta}$ 
are expressed as follows:
\begin{align*}
\mathcal{T}^{\alpha \beta} & =\operatorname{Tr}\left[\gamma^\alpha\left(\slashed{k}^\prime+m_{\nu_i}\right) \gamma^\beta\left(1-\gamma^5\right) u\left(k, r\right) \bar{u}\left(k, r\right)\left(1+\gamma^5\right)\right] \\
&= 4\left[- g^{\alpha\beta} (k - m_{\nu_i} S) \cdot k^\prime +k^{\prime\alpha}(k^\beta - m_{\nu_i} S^\beta)+ k^{\prime\beta}(k^\alpha - m_{\nu_i}S^\alpha) - i(k_\rho - m_{\nu_i} S_\rho)k^\prime_\sigma \epsilon^{\alpha\beta\rho\sigma}\right], \label{eq:19}\atag\\
\overline{\mathcal{T}}^{\alpha \beta} & =\operatorname{Tr}\left[\gamma^\alpha v\left(k, r\right) \bar{v}\left(k, r\right) \gamma^\beta\left(1-\gamma^5\right)\left(\slashed{k}^\prime-m_{\nu_i}\right)\left(1+\gamma^5\right)\right]\\
&= 4\left[ - g^{\alpha\beta} (k+m_{\nu_i} S) \cdot k^\prime +k^{\prime\alpha}(k^\beta + m_{\nu_i} S^\beta)+ k^{\prime\beta}(k^\alpha + m_{\nu_i}S^\alpha) + i(k_\rho + m_{\nu_i} S_\rho)k^\prime_\sigma \epsilon^{\alpha\beta\rho\sigma}\right],\atag\\
\mathcal{T}_X^{\alpha \beta} & =\operatorname{Tr}\left[\gamma^\alpha\left(\slashed{p}^\prime+m_X\right) \gamma^\beta\left(Q_V-Q_A \gamma^5\right)\left(\not \slashed{p}+m_X\right)\left(Q_V+Q_A \gamma^5\right)\right]\\
&= 4(Q_V^2 + Q_A^2)(-g^{\alpha\beta}(p \cdot p^\prime) + p^\alpha p^{\prime\beta} + p^\beta p^{\prime\alpha}) - 8i Q_V Q_A p_\rho p^\prime_\sigma \epsilon^{\alpha\beta\rho\sigma} + 4m_X^2 (Q_V^2 - Q_A^2)g^{\alpha\beta}. \label{eq:21} \atag
\end{align*}
where $k$, $k^\prime$, $p$, and $p^\prime$ are the momenta of incoming neutrino,  outgoing neutrino,  incoming nucleus $X$, and outgoing nucleus $X$, respectively (see Fig.~\ref{fig:feyn} left panel). Here $r$ specifies the spin state of the incoming neutrino. The outer products of spinors are given by
\begin{align}
     u\left(k, r\right) \bar{u}\left(k, r\right) &= \frac{1}{2}(\slashed{k}+m_{\nu_i})(1 + \gamma^5\slashed{S}),\\
     v\left(k, r\right) \bar{v}\left(k, r\right) &= \frac{1}{2}(\slashed{k}-m_{\nu_i})(1 + \gamma^5\slashed{S}),
\end{align}
where $S$ can be chosen as
\begin{align}
    S^\mu &= r\left(\frac{|\mathbf{k}|}{m_{\nu_i}}, \frac{E_\nu}{m_{\nu_i}} \frac{\mathbf{k}}{|\mathbf{k}|}\right).
\end{align}
With this choice, $r$ can be interpreted as the helicity of the incoming neutrino. Relic neutrinos are ultra-relativistic at decoupling, and the helicity $r$ of all relic neutrinos coincides with their chirality, which is $-1$ in the SM, at decoupling time. Since the relic neutrinos propagate through the Universe nearly freely, we can safely assume that their helicity has not been flipped since the time of decoupling.

In Eq.~\eqref{eq:21}, 
$Q_V = A - 2Z(1 - 2\sin^2 \theta_w)$ and $Q_A = A - 2Z$ are the vectorial and axial charges of the nucleus respectively, and $\theta_w$ is the weak mixing angle, with $\sin^2\theta_w\simeq 1/4$. The coherent cross-section thus acquires a large enhancement $\propto (A-Z)^2$, which depends on the number of neutrons in the target nucleus~\cite{Freedman:1973yd}. 

Evaluation of the traces~\eqref{eq:19}-\eqref{eq:21} is done using {\tt FeynCalc}~\cite{Shtabovenko:2020gxv}: 
\begin{align*}
\left\langle\left|\mathcal{M}_N\left(\nu_{i, {\sf s}}^D\right)\right|^2\right\rangle &= 8G_F^2 (k^\prime \cdot p^\prime)((k - m_{\nu_i} S )\cdot p) \\
& = 2 G_F^2 \left(-m_{\nu _i}^2-m_X^2+s\right) \left(-2 m_{\nu _i} \left(S\cdot
   p\right)-m_{\nu _i}^2-m_X^2+s\right),\atag\label{eqn: Dirac neutrino-nucleus scattering} \\
   \left\langle\left|\mathcal{M}_N\left(\bar{\nu}_{i, {\sf s}}^D\right)\right|^2\right\rangle &= 8G_F^2(p \cdot k^\prime)((k + m_{\nu_i} S) \cdot p^\prime)\\
    & =2 G_F^2 \left(-m_{\nu _i}^2-m_X^2+s+t\right) \left(-2 m_{\nu _i} \left(S\cdot
   p^\prime\right)-m_{\nu _i}^2-m_X^2+s+t\right), \atag\label{eqn: Dirac anti-neutrino-nucleus scattering}
\end{align*}
where $s, t$, and $u$ are the usual Mandelstam variables defined as
\begin{align}
    s &= (k + p)^2 = (k^\prime + p^\prime)^2,\\
    t &= (k - k^\prime)^2 = (p^\prime - p)^2,\\
    u &= (k - p^\prime)^2 = (k^\prime - p)^2.
\end{align}
In Eqs.~(\ref{eqn: Dirac neutrino-nucleus scattering}-\ref{eqn: Dirac anti-neutrino-nucleus scattering}), one of the Mandelstam variables $u$ is eliminated by using the relation $s+t+u = 2m_\nu^2 + 2m_X^2$. Since $\mathbf{p}$, the three-momentum of neutrons, is isotropic in the momentum space, only the zeroth component of any product of two four-momenta involving $p$ survives. The momentum of neutrons in the final state $\mathbf{p^\prime}$ is almost isotropic, and we can also approximate $A\cdot p^\prime \simeq A^0 p^{\prime 0}$ involving any dot product of $p'$.

Similarly, for neutrino-electron scattering, after evaluating the traces, the squared matrix elements are given by~\cite{Bauer:2022lri}
\begin{align*}
   \left\langle\left|\mathcal{M}_e\left(\nu_{i, s}^D \rightarrow \nu_j^D\right)\right|^2\right\rangle &=
    G_F^2 (1-4 \left| U_{ei}\right|^2) \delta _{ij} \left[ m_e^2 (2m_{\nu_i}S \cdot (p_e - q_{\nu_j}) - 2s + t) + (s - m_{\nu_j}^2)(- 2 m_{\nu_i} S \cdot p_e - m_{\nu_i}^2 + s) + m_e^4 \right]\\
   &+8 G_F^2
   \left| U_{ei}\right|^2 \left| U_{ej}\right| {}^2 \left(-m_e^2-m_{\nu _j}^2+s\right) \left(-m_{\nu _i} \left(2
   \left(S\cdot p_e\right)+m_{\nu _i}\right)-m_e^2+s\right) \atag \\
    \left\langle\left|\mathcal{M}_e\left(\bar{\nu}_{i, s}^D \rightarrow \bar{\nu}_j^D\right)\right|^2\right\rangle
    & = G_F^2 (1-4 \left| U_{ei}\right|^2) \delta _{ij} \left[ - m_e^2 (-2m_{\nu_i} (S \cdot q_e + S \cdot q_{\nu_j} + m_{\nu_i}) + 2 m_{\nu_j}^2 + 2s + t)\right. \\
    &\left.+ (s + t + m_{\nu_j}^2 - 2 m_{\nu_i}^2)(- 2 m_{\nu_i} S \cdot q_e - m_{\nu_i}^2 + s + t) + m_e^4 \right]\\
   &+8 G_F^2
   \left| U_{ei}\right|^2 \left| U_{ej}\right| {}^2 \left(-m_e^2+m_{\nu _j}^2 - 2 m_{\nu_i}^2 + s + t\right) \left(-m_e^2 - 2 m_{\nu_i} (S \cdot q_e) - 2 m_{\nu_i}^2 + s + t\right)\atag \, .
\end{align*}

\indent
The thermally averaged cross-section of the interaction $\nu(k^\mu,{\sf s}) + X(p^\mu, r) \to \nu(k^{\prime\mu}, q^\prime) + X(p^{\prime\mu}, r^\prime)$ of a particle $X$ in the NS with incoming neutrinos with a helicity {\sf s} is given by
\begin{align*}
    \label{eqn: cross-section 2}
    \left<\sigma\right> &= \frac{g_X}{n_X}\int \frac{d^3 p}{(2 \pi)^3} f_X(E_X) \frac{1}{4\sqrt{(k \cdot p)^2 - m_\nu^2 m_X^2}} \int \frac{d^3 k^{\prime}}{2 E^{\prime}_\nu (2 \pi)^3} \int \frac{d^3 p^{\prime}}{2 E^{\prime}_X (2 \pi)^3} \\
    &\ \ \ \ \ \times N_C \left<|\mathcal{M}|^2\right> (1-f_\nu(E_\nu^\prime))(1 \pm f_X(E_X^\prime))(2 \pi)^4 \delta^{(4)}\left(k_\mu+p_\mu-k_\mu^{\prime}-p_\mu^{\prime}\right). \atag
\end{align*}
Here  $\left<|\mathcal{M}|^2\right>$ is the scattering amplitude averaged over the helicity states of the incoming neutrino and summed over the helicity states of outgoing particles. $g_X$ is the number of degrees of freedom of $X$, $n_X$ its number density, $r,q^\prime,r^\prime$ are indices that specify spin states, $k^\mu, p^\mu, k^{\prime\mu},$ and $p^{\prime\mu}$ are four-momenta, and $N_C$ is the coherent enhancement factor [cf.~Eq.~\eqref{eq:3}]. $f_\nu$ and $f_X$ are the distribution functions for the neutrino and $X$, respectively. The factor of $1 \pm f_X(E_X^\prime)$ describes the Bose enhancement or the Pauli blocking effect depending if a particle $X$ is a boson or a fermion. Since neutron stars mostly contain neutrons, we will consider $X$ as neutrons in the rest of our calculation, with $g_n=2$.

We will perform the integral in Eq.~\eqref{eqn: cross-section 2} analytically as much as possible following Refs.~\cite{Reddy:1997yr,Bell:2020lmm}.  
The integration over $\mathbf{p^\prime}$ removes the three-dimensional delta function:
\begin{align*}
    \left<\sigma\right> &= \frac{g_n}{n_n\left<v_\mathrm{rel}\right>} \int \frac{d^3 k^{\prime}}{(2 \pi)^3} \frac{1}{(2E_\nu)(2 E^{\prime}_\nu)}\\
    &\times \int \frac{d^3 p}{(2 \pi)^3} N_C \frac{\left<|\mathcal{M}|^2\right>(s,t)}{(2 E_n)(2 E^{\prime}_n)}
     (E^\prime_n - E_n)  f_n(E_n)(1 - f_n(E_n^\prime)) (2 \pi) \delta\left(E_\nu+E_n-E^\prime_\nu-E_n^{\prime}\right).\atag
\end{align*}
Here we have treated the neutrinos inside the NS as a non-degenerate system and let the factor of $1-f_{\nu}(E_\nu^\prime) \to 1$. This is certainly valid for light relic neutrinos which eventually escape the NS, as well as for the active neutrinos produced from heavy sterile neutrino interactions in the NS.  In addition, we substitute the relation $4\sqrt{(k \cdot p)^2 - m_\nu^2 m_n^2} = (2E_\nu)(2E_n)|v_\mathrm{rel}|$ and move $|v_\mathrm{rel}|$ to the outside the integral by replacing it with its averaged value:
\begin{align}
    \left<v_\mathrm{rel}\right>(E_\nu) = \frac{\int \dfrac{d^3 p}{(2 \pi)^3} f_n(E_n) \dfrac{4\sqrt{(k \cdot p)^2 - m_\nu^2 m_n^2}}{(2E_\nu)(2E_n)}}{\int \dfrac{d^3 p}{(2 \pi)^3} f_n(E_n)}.
\end{align}
To proceed further, we write $d^3 p = p E_n d E_n d(\cos\theta_n)d\phi_n$ by defining $\theta_n$ as an angle between $\mathbf{p}$ and $\mathbf{q} = \mathbf{k} - \mathbf{k^\prime} = \mathbf{p^\prime} - \mathbf{p}$ and $\phi_n$ as that between $\mathbf{k} - (\mathbf{k}\cdot\tilde{\mathbf{q}})\tilde{\mathbf{q}}$ and $\mathbf{p} - (\mathbf{p}\cdot\tilde{\mathbf{q}})\tilde{\mathbf{q}}$ where $\tilde{\mathbf{q}} \equiv \mathbf{q}/|\mathbf{q}|$ is an unit vector parallel to $\mathbf{q}$. Letting $\Delta E$ be the energy transfer $\Delta E \equiv E_n^\prime - E_n = E_\nu - E_\nu^\prime$, we obtain 
\begin{align*}
    \left<\sigma\right> &= \frac{g_n}{n_n\left<v_\mathrm{rel}\right>}\int \frac{d^3 k^{\prime}}{(2 \pi)^3} \frac{1}{(2E_\nu)(2 E^{\prime}_\nu)} N_C \\
    &\times  \frac{1}{16\pi^2} \int \frac{p}{ E^{\prime}_n}d E_nf_n(E_n)(1 - f_n(E_n^\prime))\int d \phi_n \int^1_{-1} d \cos\theta_n \left<|\mathcal{M}|^2\right>(s,t) \Delta E  \delta\left(E_\nu+E_n-E^\prime_\nu-E_n^{\prime}\right).\atag
\end{align*}
We then next integrate over $\cos\theta_n$. The delta function fixes the value of $\cos\theta_n$:
\begin{align}
    \cos\theta_n = \frac{(\Delta E)^2 + 2E_n\Delta E - q^2}{2q\sqrt{E_n^2 - m_n^2}}.
\end{align}
As a function of $q, \Delta E$, and $\cos\theta_n$, the energy of incoming neutron $E_n$ is expressed as:
\begin{align}
    E_n = \frac{-2\Delta E\left[q^2 - (\Delta E)^2\right] \pm \sqrt{4(\Delta E)^2 \left[q^2 - (\Delta E)^2\right]^2 - 4\left[(\Delta E)^2 - q^2 \cos^2\theta_n \right](\left[q^2 - (\Delta E)^2\right]^2 + 4 q^2 m_n^2 \cos^2\theta_n)}}{4\left[ (\Delta E)^2 - q^2 \cos^2\theta_n \right]}, 
\end{align}
which has a minimum at $\cos\theta_n = \pm 1$:
\begin{align}
    E_{\mathrm{min}} = - \frac{\Delta E}{2} + \sqrt{\left( m_n + \frac{\Delta E}{2} \right)^2 + \left( \frac{\sqrt{q^2 - (\Delta E)^2}}{2} - \frac{m_n \Delta E}{\sqrt{q^2 - (\Delta E)^2}} \right)^2}.
\end{align}
and diverges to infinity at $\cos^2\theta_n = (\Delta E)^2/q^2$. The integration over $\cos\theta_n$ leaves a factor of $pq/E^\prime_n$ because $E_n^\prime = \sqrt{E_n^2 + q^2 + 2pq\cos\theta_n}$. Therefore, 
\begin{align*}
    \left<\sigma\right> &= \frac{g_n}{64 \pi^2 n_n\left<v_\mathrm{rel}\right>} \int \frac{d^3 k}{(2 \pi)^3}\int \frac{d^3 k^{\prime}}{(2 \pi)^3} N_C \frac{\Delta E}{E_\nu E^{\prime}_\nu}\frac{1}{q} \int^\infty_{E_{\mathrm{min}}}d E_n f_n(E_n)(1 - f_n(E_n^\prime))\int d \phi_n \left<|\mathcal{M}|^2\right>(s,t). \atag
\end{align*}
To find a formula of the center-of-mass energy $s = m_\nu^2 + m_n^2 + 2E_\nu E_n - 2 \mathbf{k} \cdot \mathbf{p}$ as a function of $\phi_n, q, \Delta E$, we choose the z-axis parallel to the vector $\mathbf{q}$. In this coordinate system,
\begin{align}
    \mathbf{q} &= (0,0,q),\\
    \mathbf{k} &= (k \sin\tilde{\theta}_n, 0, k \cos\tilde{\theta}_n), \\
    \mathbf{p} &= (p\sin\theta_n \cos\phi_n, p\sin\theta_n \sin\phi_n, p\cos\theta_n),
\end{align}
where $\tilde{\theta}_n$ is the angle between $\mathbf{k}$ and $\mathbf{q}$. Since
\begin{align}
    k \cos\tilde{\theta}_n &= \frac{\mathbf{k}\cdot\mathbf{q}}{|\mathbf{q}|} = - \frac{(\Delta E)^2 - q^2 - 2E_\nu \Delta E}{2q}, \\
    p\cos\theta_n &= \frac{\mathbf{p}\cdot \mathbf{q}}{|\mathbf{q}|} = - \frac{q^2 - (\Delta E)^2 - 2E_n \Delta E}{2q}, 
\end{align}
we obtain
\begin{align*}
    s &= m_\nu^2 + m_n^2 + 2E_\nu E_n - 2 \mathbf{k} \cdot \mathbf{p}  \\
    &= m_\nu^2 + m_n^2 + 2E_\nu E_n - \frac{(q^2 - (\Delta E)^2 - 2 E_n \Delta E)((\Delta E)^2 - q^2 - 2 E_\nu\Delta E)}{2q^2} \\
    &- 2\sqrt{E_n^2 - m_n^2 - \frac{(q^2 - (\Delta E)^2 - 2E_n \Delta E)^2}{4q^2}}\sqrt{E_\nu^2 - m_\nu^2 - \frac{(q^2 - (\Delta E)^2 + 2 E_\nu \Delta E)^2}{4q^2}}\cos\phi_n. \atag
\end{align*}
The $\phi_n$ integral can be evaluated analytically and results in a function of $t, E_\nu, E_n, q$, and $\Delta E$:
\begin{align}
    \int d \phi_n \left<|\mathcal{M}|^2\right>(s,t) \equiv A(t, E_\nu, E_n, q, \Delta E).
\end{align}
We next evaluate the integration over $E_n$:
\begin{align}
    \left<\sigma\right> &= \frac{g_n}{64\pi^2 n_n\left<v_\mathrm{rel}\right>} \int \frac{d^3 k}{(2 \pi)^3}\int \frac{d^3 k^{\prime}}{(2 \pi)^3}N_C \frac{\Delta E}{E_\nu E^{\prime}_\nu}\frac{1}{q} \int^\infty_{E_{\mathrm{min}}}d E_n f_n(E_n)(1 - f_n(E_n^\prime))A(t, E_\nu, E_n, q, \Delta E), \nonumber \\
    &\equiv \frac{g_n}{64 \pi ^2 n_n\left<v_\mathrm{rel}\right>} \int \frac{d^3 k}{(2 \pi)^3}\int \frac{d^3 k^{\prime}}{(2 \pi)^3} N_C \frac{\Delta E}{E_\nu E^{\prime}_\nu}\frac{1}{q} B(t, E_\nu, q, \Delta E).  
\end{align}
Next, we choose the spherical coordinates for $k^\prime$-space and choose the angle $\theta_\nu^\prime$ as the angle between $\mathbf{k}$ and $\mathbf{k^\prime}$. Using the relations
\begin{align}
    \label{eqn: coordinate transformations}
    t &= -q^2 + (\Delta E)^2, \\
    \Delta E &= E_\nu - \sqrt{k^{\prime 2} + m_\nu^2}, \\
    q^2 &= k^2 + k^{\prime 2} - 2k k^\prime \cos\theta_\nu^\prime, 
\end{align}
we perform coordinate transformations from $k^\prime$ and $\theta_\nu^\prime$ to $\Delta E$ and $t$. We thus obtain 
\begin{align}
    \left<\sigma\right>  
    &= \frac{g_n}{1024\pi^4} \frac{1}{n_n \left<v_\mathrm{rel}\right>} \frac{1}{k E_\nu} \int d(\Delta E) \Delta E  \int d t \frac{B(t, E_\nu, \Delta E)}{\sqrt{(\Delta E)^2 - t}} N_C. \atag
\end{align}
To get the equation above, we used $d (\Delta E) = - k^\prime d k^\prime/E_\nu^\prime$, $2qd q = -2k k^\prime d(\cos\theta_\nu^\prime)$, and $d t = -2q d q$. The region of integration is determined so that $|\cos\theta_\nu^\prime| \leq 1$. From Eq.~\eqref{eqn: coordinate transformations}, we find
\begin{align}
    t = 2\left[ E_\nu(\Delta E - E_\nu) + m_\nu^2 + k \sqrt{(E_\nu - \Delta E)^2 - m_\nu^2}\cos\theta_\nu^\prime \right].
\end{align}
Therefore, the lower/upper bound of the integral $t_-/t_+$ is
\begin{align}
    t_\pm = 2\left[ E_\nu(\Delta E - E_\nu) + m_\nu^2 \pm k \sqrt{(E_\nu - \Delta E)^2 - m_\nu^2} \right].
\end{align}
In summary, the cross-section averaged over the thermal distribution of neutrons is given by
\begin{align*}
    \label{eqn: energy loss rate}
    \left<\sigma\right> &= \frac{g_n}{1024\pi^4} \frac{1}{n_n \left<v_\mathrm{rel}\right>} \frac{1}{k E_\nu} \int^{E_\nu - m_\nu}_{-\infty} d(\Delta E) \Delta E  \int^{t_+}_{t_-} d t \frac{B(t, E_\nu, \Delta E)}{\sqrt{(\Delta E)^2 - t}} N_C. \atag
\end{align*}

We note that the cross-section $\langle \sigma\rangle$ shows neutrino mass-independent behavior in the regime $T_\mathrm{NS} \ll K_{\nu,\mathrm{surface}} \ll m_n$ (which is realized for the sterile neutrino case in the main text) where the $\nu-n$ cross-section scales as $\sigma \propto E_\nu^2$ while the coherent enhancement factor and a fraction $r$ of neutrons involving a scattering behave as $N_C \propto p_\nu^{-3}$ and $r \simeq |{\bf q}|/p_{f,n} \propto p_\nu$ respectively, canceling out the $m_\nu$-dependence of the cross-section. 
\section{Average Energy Transfer}
\label{app:aven}
The average energy transfer in a neutrino-neutron scattering is given by
\begin{align}
    \left<\Delta E\right> = \frac{\sum_{r,q^\prime,r^\prime} \dfrac{1}{2E_\nu} \int\ N_C |\mathcal{M}|^2 \Delta E  f_n(E_n) (1-f_\nu(E_\nu^\prime))(1 - f_n(E_n^\prime)) (2 \pi)^4 \delta^{(4)}\left(k_\mu+p_\mu-k_\mu^{\prime}-p_\mu^{\prime}\right)}{\sum_{r,q^\prime,r^\prime} \dfrac{1}{2E_\nu} \int\ N_C |\mathcal{M}|^2 f_n(E_n) (1-f_\nu(E_\nu^\prime))(1 - f_n(E_n^\prime)) (2 \pi)^4 \delta^{(4)}\left(k_\mu+p_\mu-k_\mu^{\prime}-p_\mu^{\prime}\right)},
\end{align}
where the integral is over the three-momenta $p,p',k'$: 
\begin{align}
    \int \equiv  \int \frac{d^3 p}{2 E_n (2 \pi)^3} \int \frac{d^3 k^{\prime}}{2 E^{\prime}_\nu (2 \pi)^3} \int \frac{d^3 p^{\prime}}{2 E^{\prime}_n (2 \pi)^3}.
\end{align}
As mentioned in the main text, the incoming neutrino coherently interacts with many neutrons, which lie inside a sphere with a radius equal to the de Broglie wavelength of the momentum transfer, and the corresponding enhancement factor is given by Eq.~\eqref{eq:3}. However, this equation gives inaccurate results when the mean free path $\lambda$ is shorter than the de Broglie wavelength. Therefore, we estimate the coherent enhancement factor by assuming particles inside a sphere with a radius $\lambda$ rather than $\lambda_\nu=2\pi/|\mathbf{q}|$ interact with neutrino at the same time, that is, $N_C = 4\pi \lambda^3 n_n /3$, where $\lambda$ is determined in a self-consistent way as follows: First, we divide $N_C$ by a factor of $2$ as nearly half of the sphere will be vacant in the scattering with neutrons on the surface of NS. The interaction rate, i.e. the number of collisions that each neutrino experiences per unit time, is given by
\begin{align*}
    \label{eqn: interaction rate}
    \Gamma &= \sum_{r,q^\prime,r^\prime} \frac{1}{2E_\nu} \int \frac{d^3 p}{2 E_n (2 \pi)^3} \int \frac{d^3 k^{\prime}}{2 E^{\prime}_\nu (2 \pi)^3} \int \frac{d^3 p^{\prime}}{2 E^{\prime}_n (2 \pi)^3} \frac{4\pi}{3} r \lambda^3 n_n |\mathcal{M}|^2 f_n(E_n) (1-f_\nu(E_\nu^\prime))(1 - f_n(E_n^\prime))\\
    &\ \ \ \times (2 \pi)^4 \delta^{(4)}\left(k_\mu+p_\mu-k_\mu^{\prime}-p_\mu^{\prime}\right). \atag
\end{align*}
One can see the consistency of this relation with the definition of the mean free path $\lambda \equiv 1/n_n \left<\sigma\right>$ by using the relation $4\sqrt{(k \cdot p)^2 - m_\nu^2 m_n^2} = (2E_\nu)(2E_n)|v_\mathrm{rel}|$ and $\Gamma = \left<v_\mathrm{rel}\right>/\lambda$ where $v_\mathrm{rel}$ is the relative velocity between neutrinos and neutrons. With Eq.~\eqref{eqn: interaction rate}, we define the self-consistent mean free path $\lambda$ as:
\begin{align*}
    \label{eqn: mean free path}
    \frac{1}{\lambda^4} &= \frac{1}{\left<v_\mathrm{rel}\right>}\sum_{r,q^\prime,r^\prime} \frac{1}{2E_\nu} \int \frac{d^3 p}{2 E_n (2 \pi)^3} \int \frac{d^3 k^{\prime}}{2 E^{\prime}_\nu (2 \pi)^3}\int \frac{d^3 p^{\prime}}{2 E^{\prime}_n (2 \pi)^3} \frac{4\pi}{3} r  n_n |\mathcal{M}|^2 f_n(E_n) (1-f_\nu(E_\nu^\prime))\\
    &\ \ \ \times (1 - f_n(E_n^\prime)) (2 \pi)^4 \delta^{(4)}\left(k_\mu+p_\mu-k_\mu^{\prime}-p_\mu^{\prime}\right). \atag
\end{align*}
This equation for $\lambda$ gives the precise coherent enhancement factor. We checked that self-consistency of the mean free path needs to be considered only when relic neutrinos are very light $m_\nu \ll 0.1\si{\ eV}$. For heavier relic (or sterile) neutrinos, it suffices to compute the enhancement factor by just using the momentum transfer, $N_C = 4\pi n_n/3 \times (2\pi/|\mathbf{q}|)^3$. 

\begin{figure*}[t!]
    \centering
    \includegraphics[width=0.7\linewidth]{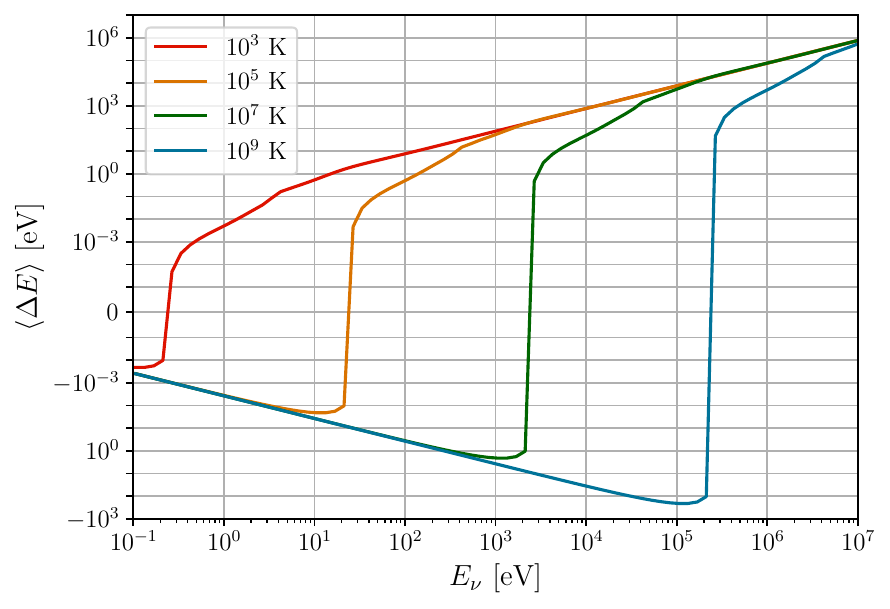}
    \caption{Average energy transfer in the first collision as a function of the incoming neutrino energy for different NS surface temperatures. Negative (Positive) $\langle \Delta E$ refers to cooling (heating) of the NS.}  
    \label{fig:aven}
\end{figure*}

We find that $\left<\Delta E\right>$ is nearly independent of NS temperature except when neutrons are more energetic than incoming relic neutrinos; see Fig.~\ref{fig:aven}.  $\left<\Delta E \right>$ increases linearly with neutrino masses, while it falls to negative values once $K_\nu$ drops below $3T_\mathrm{NS}$. This roughly defines the boundary between kinetic heating versus cooling of the NS.
\section{Number of collisions and escape probability}
\label{app:C}
The C$\nu$B on an average gains some energy from the neutrons in each interaction, which thus cools down the NS. This gain in energy increases the neutrino velocity above the escape velocity. Here we address the question how many of the neutrinos escape after ${\cal O}(1)$ interaction with the NS, thus turning the thermalization of the neutrinos with an NS ineffective. 

\begin{figure*}[t!]
    \centering
    \includegraphics[width=0.7\linewidth]{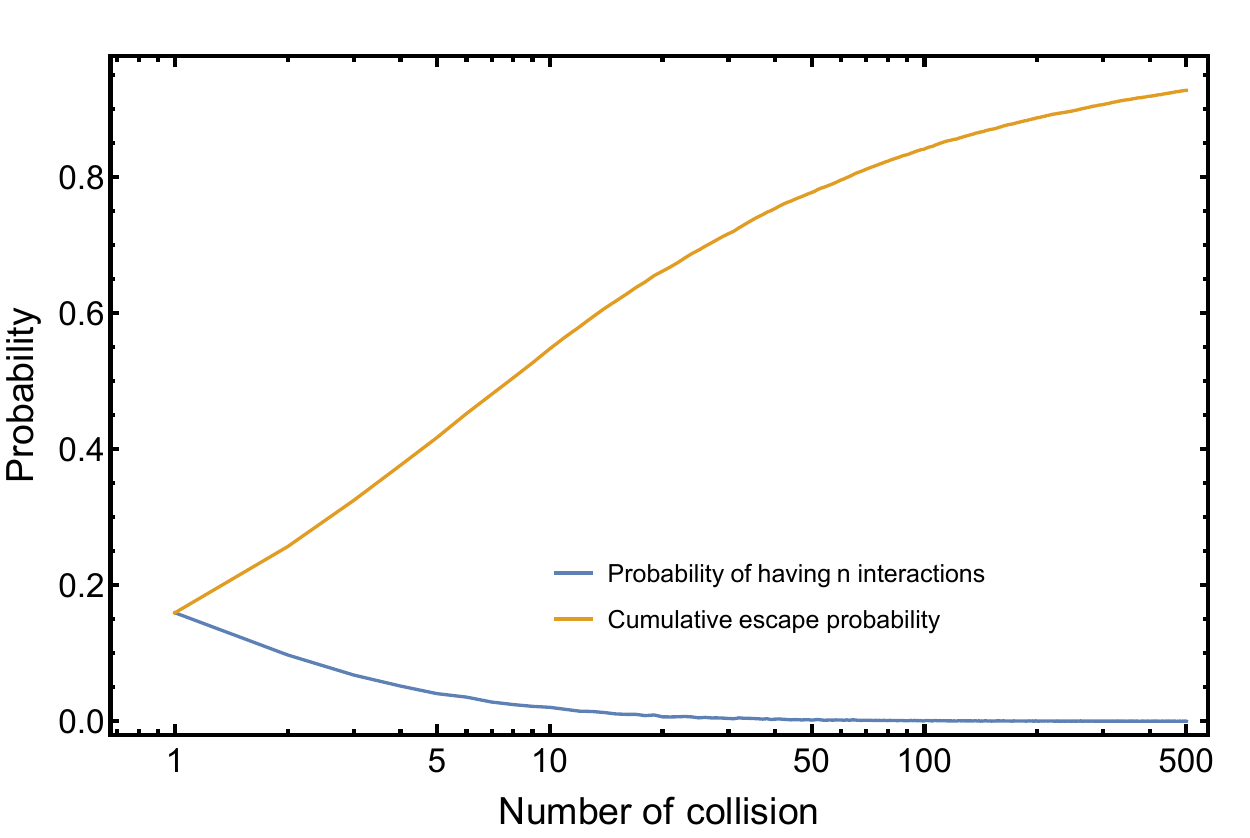}
    \caption{The escape probability of C$\nu$B as a function of the number of collisions it has inside the NS. }  
    \label{fig: Number of Collisions}
\end{figure*}

Since the neutron momentum distribution is isotropic, the neutrino random walks after each collision. For a vast region of the  parameter space, the neutrino cross section does not change parametrically with neutrino energy as the gain in single particle scattering cross section ($\propto p^2$) is cancelled by the combined effect of Pauli blocking ($\propto p$) and coherent enhancement ($\propto p^{-3}$). Assuming the neutrino cross section to be constant, we can calculate the escape probability after each collision by modeling the random walk. As shown in Fig.~\ref{fig: Number of Collisions}, the majority of the neutrinos go through multiple interactions before escape, thus making the thermalization a good approximation. Since the neutrino is not ultra-relativistic before the first interaction, the scattering cross-section slightly increases (and thus the mean free path slightly decreases) for the subsequent collision. In that case, the neutrino takes more interactions to escape. We see that the escape probability asymptotically approaches unity for many collisions, as expected.  
\section{Neutrino escape timescale}
\label{app:D}
After the neutrino thermalizes, it will go through multiple scatterings before escaping. Let us calculate the time needed on an average to move to the NS surface from the NS core. Once at the surface, the neutrino has enough kinetic energy to escape.

As in Appendix~\ref{app:C}, we can model the movement of a neutrino inside the NS as a random walk. Although probability of scattering has an angular dependence in the neutron (electron) rest frame, since the neutron (electron) velocity distribution can be assumed to be isotropic, the scattering probability is also isotropic to a good approximation. 

In random walk, the mean displacement $\sqrt{\langle r^2\rangle}$ after $N$ steps is $\propto \sqrt{N}$. If the mean free path is $\lambda$, the number of steps needed to escape the NS is then $N_{\rm max} =(R_{\rm NS}/\lambda)^2 $. Each step takes time $\delta t = \lambda + (1/m_W)$, where the first term is neutrino travel time between two collisions and the second term is the weak interaction timescale. Since weak interaction timescale is much smaller than the travel time, $\delta t\approx \lambda$, where we have used the fact that neutrino is relativistic with $\gamma=1.35$ from the gravitational clustering at the surface of the NS and becomes more relativistic as it gains energy from the NS. Then the escape time becomes 
\begin{equation}
    t_{\rm esc}=N_{\rm max}\delta t \approx \frac{R_{\rm NS}^2}{\lambda} = 300~{\rm s} \left(\frac{R_{\rm NS}}{10~{\rm km}}\right)^2\left(\frac{0.1~{\rm cm}}{\lambda}\right) \, .
    \label{eq:D1}
\end{equation}
where for the mean free path, we have used one interaction length which is given by $\lambda\sim 1/(n_n \langle\sigma\rangle)$, with $\langle\sigma\rangle$ given by Eq.~\eqref{Eq: sigma_total}. From Eq.~\eqref{eq:D1}, we see that $t_{\rm esc}\ll t_{\rm NS}\simeq 10^9$ yr. This justifies the thermalization argument for the neutrinos inside the NS. 

\bibliographystyle{utcaps_mod}
\bibliography{bibliography}

\providecommand{\href}[2]{#2}\begingroup\raggedright\begin{thebibliography}{100}

\bibitem{Weinberg:1962zza}
S.~Weinberg, ``{Universal Neutrino Degeneracy},'' \href{http://dx.doi.org/10.1103/PhysRev.128.1457}{{\em Phys. Rev.} {\bfseries 128} (1962) 1457--1473}.

\bibitem{Pitrou:2018cgg}
C.~Pitrou, A.~Coc, J.-P. Uzan, and E.~Vangioni, ``{Precision big bang nucleosynthesis with improved Helium-4 predictions},'' \href{http://dx.doi.org/10.1016/j.physrep.2018.04.005}{{\em Phys. Rept.} {\bfseries 754} (2018) 1--66}, \href{http://arxiv.org/abs/1801.08023}{[{\ttfamily 1801.08023}]}.

\bibitem{Planck:2018vyg}
{\bfseries Planck} Collaboration, N.~Aghanim {\em et~al.}, ``{Planck 2018 results. VI. Cosmological parameters},'' \href{http://dx.doi.org/10.1051/0004-6361/201833910}{{\em Astron. Astrophys.} {\bfseries 641} (2020) A6}, \href{http://arxiv.org/abs/1807.06209}{[{\ttfamily 1807.06209}]}. [Erratum: Astron.Astrophys. 652, C4 (2021)].

\bibitem{DESI:2024mwx}
{\bfseries DESI} Collaboration, A.~G. Adame {\em et~al.}, ``{DESI 2024 VI: cosmological constraints from the measurements of baryon acoustic oscillations},'' \href{http://dx.doi.org/10.1088/1475-7516/2025/02/021}{{\em JCAP} {\bfseries 02} (2025) 021}, \href{http://arxiv.org/abs/2404.03002}{[{\ttfamily 2404.03002}]}.

\bibitem{DiValentino:2024xsv}
E.~Di~Valentino, S.~Gariazzo, and O.~Mena, ``{Neutrinos in Cosmology},'' \href{http://arxiv.org/abs/2404.19322}{[{\ttfamily 2404.19322}]}.

\bibitem{Vitagliano:2019yzm}
E.~Vitagliano, I.~Tamborra, and G.~Raffelt, ``{Grand Unified Neutrino Spectrum at Earth: Sources and Spectral Components},'' \href{http://dx.doi.org/10.1103/RevModPhys.92.045006}{{\em Rev. Mod. Phys.} {\bfseries 92} (2020) 045006}, \href{http://arxiv.org/abs/1910.11878}{[{\ttfamily 1910.11878}]}.

\bibitem{Gelmini:2004hg}
G.~B. Gelmini, ``{Prospect for relic neutrino searches},'' \href{http://dx.doi.org/10.1088/0031-8949/2005/T121/019}{{\em Phys. Scripta T} {\bfseries 121} (2005) 131--136}, \href{http://arxiv.org/abs/hep-ph/0412305}{[{\ttfamily hep-ph/0412305}]}.

\bibitem{Yanagisawa:2014}
C.~{Yanagisawa}, ``{Looking for Cosmic Neutrino Background},'' \href{http://dx.doi.org/10.3389/fphy.2014.00030}{{\em Frontiers in Physics} {\bfseries 2} (2014) 30}.

\bibitem{Vogel:2015vfa}
P.~Vogel, ``{How difficult it would be to detect cosmic neutrino background?},'' \href{http://dx.doi.org/10.1063/1.4915587}{{\em AIP Conf. Proc.} {\bfseries 1666} no.~1, (2015) 140003}.

\bibitem{PTOLEMY:2018jst}
{\bfseries PTOLEMY} Collaboration, E.~Baracchini {\em et~al.}, ``{PTOLEMY: A Proposal for Thermal Relic Detection of Massive Neutrinos and Directional Detection of MeV Dark Matter},'' \href{http://arxiv.org/abs/1808.01892}{[{\ttfamily 1808.01892}]}.

\bibitem{Freedman:1973yd}
D.~Z. Freedman, ``{Coherent Neutrino Nucleus Scattering as a Probe of the Weak Neutral Current},'' \href{http://dx.doi.org/10.1103/PhysRevD.9.1389}{{\em Phys. Rev. D} {\bfseries 9} (1974) 1389--1392}.

\bibitem{Dixit:1983aj}
V.~V. Dixit and J.~Lodenquai, ``{ON NEUTRON STAR COOLING BY RELIC NEUTRINOS},'' \href{http://dx.doi.org/10.1007/BF02789550}{{\em Lett. Nuovo Cim.} {\bfseries 38} (1983) 174}.

\bibitem{Katrin:2024tvg}
{\bfseries Katrin} Collaboration, M.~Aker {\em et~al.}, ``{Direct neutrino-mass measurement based on 259 days of KATRIN data},'' \href{http://arxiv.org/abs/2406.13516}{[{\ttfamily 2406.13516}]}.

\bibitem{Drewes:2016upu}
M.~Drewes {\em et~al.}, ``{A White Paper on keV Sterile Neutrino Dark Matter},'' \href{http://dx.doi.org/10.1088/1475-7516/2017/01/025}{{\em JCAP} {\bfseries 01} (2017) 025}, \href{http://arxiv.org/abs/1602.04816}{[{\ttfamily 1602.04816}]}.

\bibitem{Gardner:2006ky}
J.~P. Gardner {\em et~al.}, ``{The James Webb Space Telescope},'' \href{http://dx.doi.org/10.1007/s11214-006-8315-7}{{\em Space Sci. Rev.} {\bfseries 123} (2006) 485}, \href{http://arxiv.org/abs/astro-ph/0606175}{[{\ttfamily astro-ph/0606175}]}.

\bibitem{2018sf2a.conf....3N}
B.~{Neichel}, D.~{Mouillet}, E.~{Gendron}, C.~{Correia}, J.~F. {Sauvage}, and T.~{Fusco}, \href{http://dx.doi.org/10.48550/arXiv.1812.06639}{``{Overview of the European Extremely Large Telescope and its instrument suite},''} in {\em SF2A-2018: Proceedings of the Annual meeting of the French Society of Astronomy and Astrophysics}, P.~{Di Matteo}, F.~{Billebaud}, F.~{Herpin}, N.~{Lagarde}, J.~B. {Marquette}, A.~{Robin}, and O.~{Venot}, eds., p.~Di.
\newblock Dec., 2018.
\newblock \href{http://arxiv.org/abs/1812.06639}{[{\ttfamily 1812.06639}]}.

\bibitem{TMT:2015pvw}
{\bfseries TMT International Science Development Teams \& TMT Science Advisory Committee} Collaboration, W.~Skidmore {\em et~al.}, ``{Thirty Meter Telescope Detailed Science Case: 2015},'' \href{http://dx.doi.org/10.1088/1674-4527/15/12/001}{{\em Res. Astron. Astrophys.} {\bfseries 15} no.~12, (2015) 1945--2140}, \href{http://arxiv.org/abs/1505.01195}{[{\ttfamily 1505.01195}]}.

\bibitem{Sawyer:1978qe}
R.~F. Sawyer and A.~Soni, ``{Transport of neutrinos in hot neutron star matter},'' \href{http://dx.doi.org/10.1086/157146}{{\em Astrophys. J.} {\bfseries 230} (1979) 859--869}.

\bibitem{Yakovlev:2004yr}
D.~G. Yakovlev, O.~Y. Gnedin, M.~E. Gusakov, A.~D. Kaminker, K.~P. Levenfish, and A.~Y. Potekhin, ``{Neutron star cooling},'' \href{http://dx.doi.org/10.1016/j.nuclphysa.2005.02.061}{{\em Nucl. Phys. A} {\bfseries 752} (2005) 590--599}, \href{http://arxiv.org/abs/astro-ph/0409751}{[{\ttfamily astro-ph/0409751}]}.

\bibitem{Page:2004fy}
D.~Page, J.~M. Lattimer, M.~Prakash, and A.~W. Steiner, ``{Minimal cooling of neutron stars: A New paradigm},'' \href{http://dx.doi.org/10.1086/424844}{{\em Astrophys. J. Suppl.} {\bfseries 155} (2004) 623--650}, \href{http://arxiv.org/abs/astro-ph/0403657}{[{\ttfamily astro-ph/0403657}]}.

\bibitem{Page:2005fq}
D.~Page, U.~Geppert, and F.~Weber, ``{The Cooling of compact stars},'' \href{http://dx.doi.org/10.1016/j.nuclphysa.2005.09.019}{{\em Nucl. Phys. A} {\bfseries 777} (2006) 497--530}, \href{http://arxiv.org/abs/astro-ph/0508056}{[{\ttfamily astro-ph/0508056}]}.

\bibitem{Ofengeim:2017cum}
D.~D. Ofengeim and D.~G. Yakovlev, ``{Analytic description of neutron star cooling},'' \href{http://dx.doi.org/10.1093/mnras/stx366}{{\em Mon. Not. Roy. Astron. Soc.} {\bfseries 467} no.~3, (2017) 3598--3603}.

\bibitem{Yakovlev:2010ed}
D.~G. Yakovlev, W.~C.~G. Ho, P.~S. Shternin, C.~O. Heinke, and A.~Y. Potekhin, ``{Cooling rates of neutron stars and the young neutron star in the Cassiopeia A supernova remnant},'' \href{http://dx.doi.org/10.1111/j.1365-2966.2010.17827.x}{{\em Mon. Not. Roy. Astron. Soc.} {\bfseries 411} (2011) 1977--1988}, \href{http://arxiv.org/abs/1010.1154}{[{\ttfamily 1010.1154}]}.

\bibitem{1994Sci...264..538W}
A.~{Wolszczan}, ``{Confirmation of Earth-Mass Planets Orbiting the Millisecond Pulsar PSR B1257+12},'' \href{http://dx.doi.org/10.1126/science.264.5158.538}{{\em Science} {\bfseries 264} no.~5158, (Apr., 1994) 538--542}.

\bibitem{Lorimer:2008se}
D.~R. Lorimer, ``{Binary and Millisecond Pulsars},'' \href{http://dx.doi.org/10.12942/lrr-2008-8}{{\em Living Rev. Rel.} {\bfseries 11} (2008) 8}, \href{http://arxiv.org/abs/0811.0762}{[{\ttfamily 0811.0762}]}.

\bibitem{Lattimer:2015nhk}
J.~M. Lattimer and M.~Prakash, ``{The Equation of State of Hot, Dense Matter and Neutron Stars},'' \href{http://dx.doi.org/10.1016/j.physrep.2015.12.005}{{\em Phys. Rept.} {\bfseries 621} (2016) 127--164}, \href{http://arxiv.org/abs/1512.07820}{[{\ttfamily 1512.07820}]}.

\bibitem{Guillot:2019ugf}
S.~Guillot, G.~G. Pavlov, C.~Reyes, A.~Reisenegger, L.~Rodriguez, B.~Rangelov, and O.~Kargaltsev, ``{Hubble Space Telescope Nondetection of PSR J2144\textendash{}3933: The Coldest Known Neutron Star},'' \href{http://dx.doi.org/10.3847/1538-4357/ab0f38}{{\em Astrophys. J.} {\bfseries 874} no.~2, (2019) 175}, \href{http://arxiv.org/abs/1901.07998}{[{\ttfamily 1901.07998}]}.

\bibitem{Ofek:2009wt}
E.~O. Ofek, ``{Space and velocity distributions of Galactic isolated old Neutron stars},'' \href{http://dx.doi.org/10.1086/605389}{{\em Publ. Astron. Soc. Pac.} {\bfseries 121} (2009) 814}, \href{http://arxiv.org/abs/0910.3684}{[{\ttfamily 0910.3684}]}.

\bibitem{Sartore:2009wn}
N.~Sartore, E.~Ripamonti, A.~Treves, and R.~Turolla, ``{Galactic neutron stars I. Space and velocity distributions in the disk and in the halo},'' \href{http://dx.doi.org/10.1051/0004-6361/200912222}{{\em Astron. Astrophys.} {\bfseries 510} (2010) A23}, \href{http://arxiv.org/abs/0908.3182}{[{\ttfamily 0908.3182}]}.

\bibitem{2016RAA....16..101T}
A.~{Taani}, ``{Systematic comparison of initial velocities for neutron stars in different models},'' \href{http://dx.doi.org/10.1088/1674-4527/16/7/101}{{\em Research in Astronomy and Astrophysics} {\bfseries 16} no.~7, (July, 2016) 101}.

\bibitem{nufit}
\url{http://www.nu-fit.org}.

\bibitem{Formaggio:2012cpf}
J.~A. Formaggio and G.~P. Zeller, ``{From eV to EeV: Neutrino Cross Sections Across Energy Scales},'' \href{http://dx.doi.org/10.1103/RevModPhys.84.1307}{{\em Rev. Mod. Phys.} {\bfseries 84} (2012) 1307--1341}, \href{http://arxiv.org/abs/1305.7513}{[{\ttfamily 1305.7513}]}.

\bibitem{COHERENT:2017ipa}
{\bfseries COHERENT} Collaboration, D.~Akimov {\em et~al.}, ``{Observation of Coherent Elastic Neutrino-Nucleus Scattering},'' \href{http://dx.doi.org/10.1126/science.aao0990}{{\em Science} {\bfseries 357} no.~6356, (2017) 1123--1126}, \href{http://arxiv.org/abs/1708.01294}{[{\ttfamily 1708.01294}]}.

\bibitem{COHERENT:2020iec}
{\bfseries COHERENT} Collaboration, D.~Akimov {\em et~al.}, ``{First Measurement of Coherent Elastic Neutrino-Nucleus Scattering on Argon},'' \href{http://dx.doi.org/10.1103/PhysRevLett.126.012002}{{\em Phys. Rev. Lett.} {\bfseries 126} no.~1, (2021) 012002}, \href{http://arxiv.org/abs/2003.10630}{[{\ttfamily 2003.10630}]}.

\bibitem{Adamski:2024yqt}
S.~Adamski {\em et~al.}, ``{First detection of coherent elastic neutrino-nucleus scattering on germanium},'' \href{http://arxiv.org/abs/2406.13806}{[{\ttfamily 2406.13806}]}.

\bibitem{Bell:2019pyc}
N.~F. Bell, G.~Busoni, and S.~Robles, ``{Capture of Leptophilic Dark Matter in Neutron Stars},'' \href{http://dx.doi.org/10.1088/1475-7516/2019/06/054}{{\em JCAP} {\bfseries 06} (2019) 054}, \href{http://arxiv.org/abs/1904.09803}{[{\ttfamily 1904.09803}]}.

\bibitem{Pearson:2018tkr}
J.~M. Pearson, N.~Chamel, A.~Y. Potekhin, A.~F. Fantina, C.~Ducoin, A.~K. Dutta, and S.~Goriely, ``{Unified equations of state for cold non-accreting neutron stars with Brussels\textendash{}Montreal functionals \textendash{} I. Role of symmetry energy},'' \href{http://dx.doi.org/10.1093/mnras/sty2413}{{\em Mon. Not. Roy. Astron. Soc.} {\bfseries 481} no.~3, (2018) 2994--3026}, \href{http://arxiv.org/abs/1903.04981}{[{\ttfamily 1903.04981}]}. [Erratum: Mon.Not.Roy.Astron.Soc. 486, 768 (2019)].

\bibitem{Goldman:1989nd}
I.~Goldman and S.~Nussinov, ``{Weakly Interacting Massive Particles and Neutron Stars},'' \href{http://dx.doi.org/10.1103/PhysRevD.40.3221}{{\em Phys. Rev. D} {\bfseries 40} (1989) 3221--3230}.

\bibitem{Arvanitaki:2022oby}
A.~Arvanitaki and S.~Dimopoulos, ``{Cosmic neutrino background on the surface of the Earth},'' \href{http://dx.doi.org/10.1103/PhysRevD.108.043517}{{\em Phys. Rev. D} {\bfseries 108} no.~4, (2023) 043517}, \href{http://arxiv.org/abs/2212.00036}{[{\ttfamily 2212.00036}]}.

\bibitem{Huang:2024tog}
G.-y. Huang, ``{Neutrino-antineutrino asymmetry of C\ensuremath{\nu}B on the surface of the round Earth},'' \href{http://dx.doi.org/10.1007/JHEP11(2024)153}{{\em JHEP} {\bfseries 11} (2024) 153}, \href{http://arxiv.org/abs/2401.07347}{[{\ttfamily 2401.07347}]}.

\bibitem{Kalia:2024xeq}
S.~Kalia, ``{Tunneling away the relic neutrino asymmetry},'' \href{http://dx.doi.org/10.1103/PhysRevD.110.053001}{{\em Phys. Rev. D} {\bfseries 110} no.~5, (2024) 053001}, \href{http://arxiv.org/abs/2404.11664}{[{\ttfamily 2404.11664}]}.

\bibitem{Ringwald:2004np}
A.~Ringwald and Y.~Y.~Y. Wong, ``{Gravitational clustering of relic neutrinos and implications for their detection},'' \href{http://dx.doi.org/10.1088/1475-7516/2004/12/005}{{\em JCAP} {\bfseries 12} (2004) 005}, \href{http://arxiv.org/abs/hep-ph/0408241}{[{\ttfamily hep-ph/0408241}]}.

\bibitem{Mertsch:2019qjv}
P.~Mertsch, G.~Parimbelli, P.~F. de~Salas, S.~Gariazzo, J.~Lesgourgues, and S.~Pastor, ``{Neutrino clustering in the Milky Way and beyond},'' \href{http://dx.doi.org/10.1088/1475-7516/2020/01/015}{{\em JCAP} {\bfseries 01} (2020) 015}, \href{http://arxiv.org/abs/1910.13388}{[{\ttfamily 1910.13388}]}.

\bibitem{Zimmer:2023jbb}
F.~Zimmer, C.~A. Correa, and S.~Ando, ``{Influence of local structure on relic neutrino abundances and anisotropies},'' \href{http://dx.doi.org/10.1088/1475-7516/2023/11/038}{{\em JCAP} {\bfseries 11} (2023) 038}, \href{http://arxiv.org/abs/2306.16444}{[{\ttfamily 2306.16444}]}.

\bibitem{Holm:2024zpr}
E.~B. Holm, S.~Zentarra, and I.~M. Oldengott, ``{Local clustering of relic neutrinos: comparison of kinetic field theory and the Vlasov equation},'' \href{http://dx.doi.org/10.1088/1475-7516/2024/07/050}{{\em JCAP} {\bfseries 07} (2024) 050}, \href{http://arxiv.org/abs/2404.11295}{[{\ttfamily 2404.11295}]}.

\bibitem{Baryakhtar:2017dbj}
M.~Baryakhtar, J.~Bramante, S.~W. Li, T.~Linden, and N.~Raj, ``{Dark Kinetic Heating of Neutron Stars and An Infrared Window On WIMPs, SIMPs, and Pure Higgsinos},'' \href{http://dx.doi.org/10.1103/PhysRevLett.119.131801}{{\em Phys. Rev. Lett.} {\bfseries 119} no.~13, (2017) 131801}, \href{http://arxiv.org/abs/1704.01577}{[{\ttfamily 1704.01577}]}.

\bibitem{KATRIN:2022kkv}
{\bfseries KATRIN} Collaboration, M.~Aker {\em et~al.}, ``{New Constraint on the Local Relic Neutrino Background Overdensity with the First KATRIN Data Runs},'' \href{http://dx.doi.org/10.1103/PhysRevLett.129.011806}{{\em Phys. Rev. Lett.} {\bfseries 129} no.~1, (2022) 011806}, \href{http://arxiv.org/abs/2202.04587}{[{\ttfamily 2202.04587}]}.

\bibitem{Brdar:2022kpu}
V.~Brdar, P.~S.~B. Dev, R.~Plestid, and A.~Soni, ``{A new probe of relic neutrino clustering using cosmogenic neutrinos},'' \href{http://dx.doi.org/10.1016/j.physletb.2022.137358}{{\em Phys. Lett. B} {\bfseries 833} (2022) 137358}, \href{http://arxiv.org/abs/2207.02860}{[{\ttfamily 2207.02860}]}.

\bibitem{Bauer:2022lri}
M.~Bauer and J.~D. Shergold, ``{Limits on the cosmic neutrino background},'' \href{http://dx.doi.org/10.1088/1475-7516/2023/01/003}{{\em JCAP} {\bfseries 01} (2023) 003}, \href{http://arxiv.org/abs/2207.12413}{[{\ttfamily 2207.12413}]}.

\bibitem{Tsai:2022jnv}
Y.-D. Tsai, J.~Eby, J.~Arakawa, D.~Farnocchia, and M.~S. Safronova, ``{OSIRIS-REx constraints on local dark matter and cosmic neutrino profiles},'' \href{http://dx.doi.org/10.1088/1475-7516/2024/02/029}{{\em JCAP} {\bfseries 02} (2024) 029}, \href{http://arxiv.org/abs/2210.03749}{[{\ttfamily 2210.03749}]}.

\bibitem{Ciscar-Monsalvatje:2024tvm}
M.~C\'\i{}scar-Monsalvatje, G.~Herrera, and I.~M. Shoemaker, ``{Upper limits on the cosmic neutrino background from cosmic rays},'' \href{http://dx.doi.org/10.1103/PhysRevD.110.063036}{{\em Phys. Rev. D} {\bfseries 110} no.~6, (2024) 063036}, \href{http://arxiv.org/abs/2402.00985}{[{\ttfamily 2402.00985}]}.

\bibitem{Franklin:2024enc}
J.~Franklin, I.~Martinez-Soler, Y.~F. Perez-Gonzalez, and J.~Turner, ``{Constraints on the Cosmic Neutrino Background from NGC 1068},'' \href{http://arxiv.org/abs/2404.02202}{[{\ttfamily 2404.02202}]}.

\bibitem{DeMarchi:2024zer}
A.~G. De~Marchi, A.~Granelli, J.~Nava, and F.~Sala, ``{Relic Neutrino Background from Cosmic-Ray Reservoirs},'' \href{http://arxiv.org/abs/2405.04568}{[{\ttfamily 2405.04568}]}.

\bibitem{Herrera:2024upj}
G.~Herrera, S.~Horiuchi, and X.~Qi, ``{Diffuse boosted cosmic neutrino background},'' \href{http://dx.doi.org/10.1103/PhysRevD.111.063016}{{\em Phys. Rev. D} {\bfseries 111} no.~6, (2025) 063016}, \href{http://arxiv.org/abs/2405.14946}{[{\ttfamily 2405.14946}]}.

\bibitem{Smirnov:2022sfo}
A.~Y. Smirnov and X.-J. Xu, ``{Neutrino bound states and bound systems},'' \href{http://dx.doi.org/10.1007/JHEP08(2022)170}{{\em JHEP} {\bfseries 08} (2022) 170}, \href{http://arxiv.org/abs/2201.00939}{[{\ttfamily 2201.00939}]}.

\bibitem{Fardon:2003eh}
R.~Fardon, A.~E. Nelson, and N.~Weiner, ``{Dark energy from mass varying neutrinos},'' \href{http://dx.doi.org/10.1088/1475-7516/2004/10/005}{{\em JCAP} {\bfseries 10} (2004) 005}, \href{http://arxiv.org/abs/astro-ph/0309800}{[{\ttfamily astro-ph/0309800}]}.

\bibitem{Krnjaic:2017zlz}
G.~Krnjaic, P.~A.~N. Machado, and L.~Necib, ``{Distorted neutrino oscillations from time varying cosmic fields},'' \href{http://dx.doi.org/10.1103/PhysRevD.97.075017}{{\em Phys. Rev. D} {\bfseries 97} no.~7, (2018) 075017}, \href{http://arxiv.org/abs/1705.06740}{[{\ttfamily 1705.06740}]}.

\bibitem{Lorenz:2018fzb}
C.~S. Lorenz, L.~Funcke, E.~Calabrese, and S.~Hannestad, ``{Time-varying neutrino mass from a supercooled phase transition: current cosmological constraints and impact on the $\Omega_m$-$\sigma_8$ plane},'' \href{http://dx.doi.org/10.1103/PhysRevD.99.023501}{{\em Phys. Rev. D} {\bfseries 99} no.~2, (2019) 023501}, \href{http://arxiv.org/abs/1811.01991}{[{\ttfamily 1811.01991}]}.

\bibitem{Huang:2022wmz}
G.-y. Huang, M.~Lindner, P.~Mart\'\i{}nez-Mirav\'e, and M.~Sen, ``{Cosmology-friendly time-varying neutrino masses via the sterile neutrino portal},'' \href{http://dx.doi.org/10.1103/PhysRevD.106.033004}{{\em Phys. Rev. D} {\bfseries 106} no.~3, (2022) 033004}, \href{http://arxiv.org/abs/2205.08431}{[{\ttfamily 2205.08431}]}.

\bibitem{Goertz:2024gzw}
F.~Goertz, M.~Hager, G.~Laverda, and J.~Rubio, ``{Phasing out of Darkness: From Sterile Neutrino Dark Matter to Neutrino Masses via Time-Dependent Mixing},'' \href{http://arxiv.org/abs/2407.04778}{[{\ttfamily 2407.04778}]}.

\bibitem{ParticleDataGroup:2022pth}
{\bfseries Particle Data Group} Collaboration, R.~L. Workman {\em et~al.}, ``{Review of Particle Physics},'' \href{http://dx.doi.org/10.1093/ptep/ptac097}{{\em PTEP} {\bfseries 2022} (2022) 083C01}.

\bibitem{Kouvaris:2007ay}
C.~Kouvaris, ``{WIMP Annihilation and Cooling of Neutron Stars},'' \href{http://dx.doi.org/10.1103/PhysRevD.77.023006}{{\em Phys. Rev. D} {\bfseries 77} (2008) 023006}, \href{http://arxiv.org/abs/0708.2362}{[{\ttfamily 0708.2362}]}.

\bibitem{Kouvaris:2010vv}
C.~Kouvaris and P.~Tinyakov, ``{Can Neutron stars constrain Dark Matter?},'' \href{http://dx.doi.org/10.1103/PhysRevD.82.063531}{{\em Phys. Rev. D} {\bfseries 82} (2010) 063531}, \href{http://arxiv.org/abs/1004.0586}{[{\ttfamily 1004.0586}]}.

\bibitem{deLavallaz:2010wp}
A.~de~Lavallaz and M.~Fairbairn, ``{Neutron Stars as Dark Matter Probes},'' \href{http://dx.doi.org/10.1103/PhysRevD.81.123521}{{\em Phys. Rev. D} {\bfseries 81} (2010) 123521}, \href{http://arxiv.org/abs/1004.0629}{[{\ttfamily 1004.0629}]}.

\bibitem{Bramante:2017xlb}
J.~Bramante, A.~Delgado, and A.~Martin, ``{Multiscatter stellar capture of dark matter},'' \href{http://dx.doi.org/10.1103/PhysRevD.96.063002}{{\em Phys. Rev. D} {\bfseries 96} no.~6, (2017) 063002}, \href{http://arxiv.org/abs/1703.04043}{[{\ttfamily 1703.04043}]}.

\bibitem{Raj:2017wrv}
N.~Raj, P.~Tanedo, and H.-B. Yu, ``{Neutron stars at the dark matter direct detection frontier},'' \href{http://dx.doi.org/10.1103/PhysRevD.97.043006}{{\em Phys. Rev. D} {\bfseries 97} no.~4, (2018) 043006}, \href{http://arxiv.org/abs/1707.09442}{[{\ttfamily 1707.09442}]}.

\bibitem{Bell:2018pkk}
N.~F. Bell, G.~Busoni, and S.~Robles, ``{Heating up Neutron Stars with Inelastic Dark Matter},'' \href{http://dx.doi.org/10.1088/1475-7516/2018/09/018}{{\em JCAP} {\bfseries 09} (2018) 018}, \href{http://arxiv.org/abs/1807.02840}{[{\ttfamily 1807.02840}]}.

\bibitem{Joglekar:2019vzy}
A.~Joglekar, N.~Raj, P.~Tanedo, and H.-B. Yu, ``{Relativistic capture of dark matter by electrons in neutron stars},'' \href{http://dx.doi.org/10.1016/j.physletb.2020.135767}{{\em Phys. Lett. B} {\bfseries 809} (2020) 135767}, \href{http://arxiv.org/abs/1911.13293}{[{\ttfamily 1911.13293}]}.

\bibitem{Joglekar:2020liw}
A.~Joglekar, N.~Raj, P.~Tanedo, and H.-B. Yu, ``{Dark kinetic heating of neutron stars from contact interactions with relativistic targets},'' \href{http://dx.doi.org/10.1103/PhysRevD.102.123002}{{\em Phys. Rev. D} {\bfseries 102} no.~12, (2020) 123002}, \href{http://arxiv.org/abs/2004.09539}{[{\ttfamily 2004.09539}]}.

\bibitem{Bell:2020jou}
N.~F. Bell, G.~Busoni, S.~Robles, and M.~Virgato, ``{Improved Treatment of Dark Matter Capture in Neutron Stars},'' \href{http://dx.doi.org/10.1088/1475-7516/2020/09/028}{{\em JCAP} {\bfseries 09} (2020) 028}, \href{http://arxiv.org/abs/2004.14888}{[{\ttfamily 2004.14888}]}.

\bibitem{Bell:2020lmm}
N.~F. Bell, G.~Busoni, S.~Robles, and M.~Virgato, ``{Improved Treatment of Dark Matter Capture in Neutron Stars II: Leptonic Targets},'' \href{http://dx.doi.org/10.1088/1475-7516/2021/03/086}{{\em JCAP} {\bfseries 03} (2021) 086}, \href{http://arxiv.org/abs/2010.13257}{[{\ttfamily 2010.13257}]}.

\bibitem{Chatterjee:2022dhp}
S.~Chatterjee, R.~Garani, R.~K. Jain, B.~Kanodia, M.~S.~N. Kumar, and S.~K. Vempati, ``{Faint light of old neutron stars and detectability at the James Webb Space Telescope},'' \href{http://dx.doi.org/10.1103/PhysRevD.108.L021301}{{\em Phys. Rev. D} {\bfseries 108} no.~2, (2023) L021301}, \href{http://arxiv.org/abs/2205.05048}{[{\ttfamily 2205.05048}]}.

\bibitem{Avila:2023rzj}
A.~\'Avila, E.~Giangrandi, V.~Sagun, O.~Ivanytskyi, and C.~Provid\^encia, ``{Rapid neutron star cooling triggered by dark matter},'' \href{http://dx.doi.org/10.1093/mnras/stae337}{{\em Mon. Not. Roy. Astron. Soc.} {\bfseries 528} no.~4, (2024) 6319--6328}, \href{http://arxiv.org/abs/2309.03894}{[{\ttfamily 2309.03894}]}.

\bibitem{Bell:2023ysh}
N.~F. Bell, G.~Busoni, S.~Robles, and M.~Virgato, ``{Thermalization and annihilation of dark matter in neutron stars},'' \href{http://dx.doi.org/10.1088/1475-7516/2024/04/006}{{\em JCAP} {\bfseries 04} (2024) 006}, \href{http://arxiv.org/abs/2312.11892}{[{\ttfamily 2312.11892}]}.

\bibitem{Bramante:2023djs}
J.~Bramante and N.~Raj, ``{Dark matter in compact stars},'' \href{http://dx.doi.org/10.1016/j.physrep.2023.12.001}{{\em Phys. Rept.} {\bfseries 1052} (2024) 1--48}, \href{http://arxiv.org/abs/2307.14435}{[{\ttfamily 2307.14435}]}.

\bibitem{Minkowski:1977sc}
P.~Minkowski, ``{$\mu \to e\gamma$ at a Rate of One Out of $10^{9}$ Muon Decays?},'' \href{http://dx.doi.org/10.1016/0370-2693(77)90435-X}{{\em Phys. Lett. B} {\bfseries 67} (1977) 421--428}.

\bibitem{Mohapatra:1979ia}
R.~N. Mohapatra and G.~Senjanovic, ``{Neutrino Mass and Spontaneous Parity Nonconservation},'' \href{http://dx.doi.org/10.1103/PhysRevLett.44.912}{{\em Phys. Rev. Lett.} {\bfseries 44} (1980) 912}.

\bibitem{Yanagida:1979as}
T.~Yanagida, ``{Horizontal gauge symmetry and masses of neutrinos},'' {\em Conf. Proc. C} {\bfseries 7902131} (1979) 95--99.

\bibitem{Gell-Mann:1979vob}
M.~Gell-Mann, P.~Ramond, and R.~Slansky, ``{Complex Spinors and Unified Theories},'' {\em Conf. Proc. C} {\bfseries 790927} (1979) 315--321, \href{http://arxiv.org/abs/1306.4669}{[{\ttfamily 1306.4669}]}.

\bibitem{Asaka:2005an}
T.~Asaka, S.~Blanchet, and M.~Shaposhnikov, ``{The nuMSM, dark matter and neutrino masses},'' \href{http://dx.doi.org/10.1016/j.physletb.2005.09.070}{{\em Phys. Lett. B} {\bfseries 631} (2005) 151--156}, \href{http://arxiv.org/abs/hep-ph/0503065}{[{\ttfamily hep-ph/0503065}]}.

\bibitem{Callingham:2018vcf}
T.~M. Callingham, M.~Cautun, A.~J. Deason, C.~S. Frenk, W.~Wang, F.~A. G\'omez, R.~J.~J. Grand, F.~Marinacci, R.~Pakmor, and R.~Pakmor, ``{The mass of the Milky Way from satellite dynamics},'' \href{http://dx.doi.org/10.1093/mnras/stz365}{{\em Mon. Not. Roy. Astron. Soc.} {\bfseries 484} no.~4, (2019) 5453--5467}, \href{http://arxiv.org/abs/1808.10456}{[{\ttfamily 1808.10456}]}.

\bibitem{Ando:2010ye}
S.~Ando and A.~Kusenko, ``{Interactions of keV sterile neutrinos with matter},'' \href{http://dx.doi.org/10.1103/PhysRevD.81.113006}{{\em Phys. Rev. D} {\bfseries 81} (2010) 113006}, \href{http://arxiv.org/abs/1001.5273}{[{\ttfamily 1001.5273}]}.

\bibitem{2009PASP..121..814O}
E.~O. {Ofek}, ``{Space and Velocity Distributions of Galactic Isolated Old Neutron Stars},'' \href{http://dx.doi.org/10.1086/605389}{{\em \pasp} {\bfseries 121} no.~882, (Aug., 2009) 814}, \href{http://arxiv.org/abs/0910.3684}{[{\ttfamily 0910.3684}]}.

\bibitem{Pal:1981rm}
P.~B. Pal and L.~Wolfenstein, ``{Radiative Decays of Massive Neutrinos},'' \href{http://dx.doi.org/10.1103/PhysRevD.25.766}{{\em Phys. Rev. D} {\bfseries 25} (1982) 766}.

\bibitem{Barger:1995ty}
V.~D. Barger, R.~J.~N. Phillips, and S.~Sarkar, ``{Remarks on the KARMEN anomaly},'' \href{http://dx.doi.org/10.1016/0370-2693(95)00486-5}{{\em Phys. Lett. B} {\bfseries 352} (1995) 365--371}, \href{http://arxiv.org/abs/hep-ph/9503295}{[{\ttfamily hep-ph/9503295}]}. [Erratum: Phys.Lett.B 356, 617--617 (1995)].

\bibitem{Ng:2019gch}
K.~C.~Y. Ng, B.~M. Roach, K.~Perez, J.~F. Beacom, S.~Horiuchi, R.~Krivonos, and D.~R. Wik, ``{New Constraints on Sterile Neutrino Dark Matter from $NuSTAR$ M31 Observations},'' \href{http://dx.doi.org/10.1103/PhysRevD.99.083005}{{\em Phys. Rev.} {\bfseries D99} (2019) 083005},
\href{http://arxiv.org/abs/1901.01262}{[{\ttfamily 1901.01262}]}.

\bibitem{Roach:2019ctw}
B.~M. Roach, K.~C. Ng, K.~Perez, J.~F. Beacom, S.~Horiuchi, R.~Krivonos, and D.~R. Wik, ``{NuSTAR Tests of Sterile-Neutrino Dark Matter: New Galactic Bulge Observations and Combined Impact},'' \href{http://dx.doi.org/10.1103/PhysRevD.101.103011}{{\em Phys. Rev. D} {\bfseries 101} no.~10, (2020) 103011}, \href{http://arxiv.org/abs/1908.09037}{[{\ttfamily 1908.09037}]}.

\bibitem{Laha:2020ivk}
R.~Laha, J.~B. Mu\~noz, and T.~R. Slatyer, ``{INTEGRAL constraints on primordial black holes and particle dark matter},'' \href{http://dx.doi.org/10.1103/PhysRevD.101.123514}{{\em Phys. Rev. D} {\bfseries 101} no.~12, (2020) 123514}, \href{http://arxiv.org/abs/2004.00627}{[{\ttfamily 2004.00627}]}.

\bibitem{Neronov:2015kca}
A.~Neronov and D.~Malyshev, ``{Toward a full test of the $\nu$MSM sterile neutrino dark matter model with Athena},'' \href{http://dx.doi.org/10.1103/PhysRevD.93.063518}{{\em Phys. Rev.} {\bfseries D93} no.~6, (2016) 063518},
\href{http://arxiv.org/abs/1509.02758}{[{\ttfamily 1509.02758}]}.

\bibitem{Dekker:2021bos}
A.~Dekker, E.~Peerbooms, F.~Zimmer, K.~C.~Y. Ng, and S.~Ando, ``{Searches for sterile neutrinos and axionlike particles from the Galactic halo with eROSITA},'' \href{http://dx.doi.org/10.1103/PhysRevD.104.023021}{{\em Phys. Rev. D} {\bfseries 104} no.~2, (2021) 023021}, \href{http://arxiv.org/abs/2103.13241}{[{\ttfamily 2103.13241}]}.

\bibitem{Shi:1993ee}
X.~Shi and G.~Sigl, ``{A Type II supernovae constraint on electron-neutrino - sterile-neutrino mixing},'' \href{http://dx.doi.org/10.1016/0370-2693(94)90233-X, 10.1016/0370-2693(94)91232-7}{{\em Phys. Lett.} {\bfseries B323} (1994) 360--366}, \href{http://arxiv.org/abs/hep-ph/9312247}{[{\ttfamily hep-ph/9312247}]}.
[Erratum: Phys. Lett.B324,516(1994)].

\bibitem{Vincent:2014rja}
A.~C. Vincent, E.~F. Martinez, P.~HernÃ¡ndez, M.~Lattanzi, and O.~Mena, ``{Revisiting cosmological bounds on sterile neutrinos},'' \href{http://dx.doi.org/10.1088/1475-7516/2015/04/006}{{\em JCAP} {\bfseries 1504} no.~04, (2015) 006},
\href{http://arxiv.org/abs/1408.1956}{[{\ttfamily 1408.1956}]}.

\bibitem{Bridle:2016isd}
S.~Bridle, J.~Elvin-Poole, J.~Evans, S.~Fernandez, P.~Guzowski, and S.~Soldner-Rembold, ``{A Combined View of Sterile-Neutrino Constraints from CMB and Neutrino Oscillation Measurements},'' \href{http://dx.doi.org/10.1016/j.physletb.2016.11.050}{{\em Phys. Lett.} {\bfseries B764} (2017) 322--327},
\href{http://arxiv.org/abs/1607.00032}{[{\ttfamily 1607.00032}]}.

\bibitem{Garzilli:2019qki}
A.~Garzilli, A.~Magalich, O.~Ruchayskiy, and A.~Boyarsky, ``{How to constrain warm dark matter with the Lyman-$\alpha$ forest},'' \href{http://dx.doi.org/10.1093/mnras/stab192}{{\em Mon. Not. Roy. Astron. Soc.} {\bfseries 502} no.~2, (2021) 2356--2363}, \href{http://arxiv.org/abs/1912.09397}{[{\ttfamily 1912.09397}]}.

\bibitem{Irsic:2023equ}
V.~Ir\v{s}i\v{c} {\em et~al.}, ``{Unveiling dark matter free streaming at the smallest scales with the high redshift Lyman-alpha forest},'' \href{http://dx.doi.org/10.1103/PhysRevD.109.043511}{{\em Phys. Rev. D} {\bfseries 109} no.~4, (2024) 043511}, \href{http://arxiv.org/abs/2309.04533}{[{\ttfamily 2309.04533}]}.

\bibitem{Horiuchi:2013noa}
S.~Horiuchi, P.~J. Humphrey, J.~Onorbe, K.~N. Abazajian, M.~Kaplinghat, and S.~Garrison-Kimmel, ``{Sterile neutrino dark matter bounds from galaxies of the Local Group},'' \href{http://dx.doi.org/10.1103/PhysRevD.89.025017}{{\em Phys. Rev. D} {\bfseries 89} no.~2, (2014) 025017}, \href{http://arxiv.org/abs/1311.0282}{[{\ttfamily 1311.0282}]}.

\bibitem{Dekker:2021scf}
A.~Dekker, S.~Ando, C.~A. Correa, and K.~C.~Y. Ng, ``{Warm dark matter constraints using Milky~Way satellite observations and subhalo evolution modeling},'' \href{http://dx.doi.org/10.1103/PhysRevD.106.123026}{{\em Phys. Rev. D} {\bfseries 106} no.~12, (2022) 123026}, \href{http://arxiv.org/abs/2111.13137}{[{\ttfamily 2111.13137}]}.

\bibitem{DES:2020fxi}
{\bfseries DES} Collaboration, E.~O. Nadler {\em et~al.}, ``{Milky Way Satellite Census. III. Constraints on Dark Matter Properties from Observations of Milky Way Satellite Galaxies},'' \href{http://dx.doi.org/10.1103/PhysRevLett.126.091101}{{\em Phys. Rev. Lett.} {\bfseries 126} (2021) 091101}, \href{http://arxiv.org/abs/2008.00022}{[{\ttfamily 2008.00022}]}.

\bibitem{Tremaine:1979we}
S.~Tremaine and J.~E. Gunn, ``{Dynamical Role of Light Neutral Leptons in Cosmology},'' \href{http://dx.doi.org/10.1103/PhysRevLett.42.407}{{\em Phys. Rev. Lett.} {\bfseries 42} (1979) 407--410}.

\bibitem{Dodelson:1993je}
S.~Dodelson and L.~M. Widrow, ``{Sterile-neutrinos as dark matter},'' \href{http://dx.doi.org/10.1103/PhysRevLett.72.17}{{\em Phys. Rev. Lett.} {\bfseries 72} (1994) 17--20}, \href{http://arxiv.org/abs/hep-ph/9303287}{[{\ttfamily hep-ph/9303287}]}.

\bibitem{Shi:1998km}
X.-D. Shi and G.~M. Fuller, ``{A New dark matter candidate: Nonthermal sterile neutrinos},'' \href{http://dx.doi.org/10.1103/PhysRevLett.82.2832}{{\em Phys. Rev. Lett.} {\bfseries 82} (1999) 2832--2835}, \href{http://arxiv.org/abs/astro-ph/9810076}{[{\ttfamily astro-ph/9810076}]}.

\bibitem{An:2023mkf}
R.~An, V.~Gluscevic, E.~O. Nadler, and Y.~Zhang, ``{Can Neutrino Self-interactions Save Sterile Neutrino Dark Matter?},'' \href{http://dx.doi.org/10.3847/2041-8213/acf049}{{\em Astrophys. J. Lett.} {\bfseries 954} no.~1, (2023) L18}, \href{http://arxiv.org/abs/2301.08299}{[{\ttfamily 2301.08299}]}.

\bibitem{Astros:2023xhe}
M.~D. Astros and S.~Vogl, ``{Boosting the production of sterile neutrino dark matter with self-interactions},'' \href{http://dx.doi.org/10.1007/JHEP03(2024)032}{{\em JHEP} {\bfseries 03} (2024) 032}, \href{http://arxiv.org/abs/2307.15565}{[{\ttfamily 2307.15565}]}.

\bibitem{Schneider:2016uqi}
A.~Schneider, ``{Astrophysical constraints on resonantly produced sterile neutrino dark matter},'' \href{http://dx.doi.org/10.1088/1475-7516/2016/04/059}{{\em JCAP} {\bfseries 04} (2016) 059}, \href{http://arxiv.org/abs/1601.07553}{[{\ttfamily 1601.07553}]}.

\bibitem{Atre:2009rg}
A.~Atre, T.~Han, S.~Pascoli, and B.~Zhang, ``{The Search for Heavy Majorana Neutrinos},'' \href{http://dx.doi.org/10.1088/1126-6708/2009/05/030}{{\em JHEP} {\bfseries 0905} (2009) 030}, \href{http://arxiv.org/abs/0901.3589}{[{\ttfamily 0901.3589}]}.

\bibitem{Bolton:2019pcu}
P.~D. Bolton, F.~F. Deppisch, and P.~S. Bhupal~Dev, ``{Neutrinoless double beta decay versus other probes of heavy sterile neutrinos},'' \href{http://dx.doi.org/10.1007/JHEP03(2020)170}{{\em JHEP} {\bfseries 03} (2020) 170}, \href{http://arxiv.org/abs/1912.03058}{[{\ttfamily 1912.03058}]}.

\bibitem{KATRIN:2018oow}
{\bfseries KATRIN} Collaboration, S.~Mertens {\em et~al.}, ``{A novel detector system for KATRIN to search for keV-scale sterile neutrinos},'' \href{http://dx.doi.org/10.1088/1361-6471/ab12fe}{{\em J. Phys. G} {\bfseries 46} no.~6, (2019) 065203}, \href{http://arxiv.org/abs/1810.06711}{[{\ttfamily 1810.06711}]}.

\bibitem{Martoff:2021vxp}
C.~J. Martoff {\em et~al.}, ``{HUNTER: precision massive-neutrino search based on a laser cooled atomic source},'' \href{http://dx.doi.org/10.1088/2058-9565/abdb9b}{{\em Quantum Sci. Technol.} {\bfseries 6} no.~2, (2021) 024008}.

\bibitem{Arguelles:2016uwb}
C.~A. ArgÃ¼elles, V.~Brdar, and J.~Kopp, ``{Production of keV Sterile Neutrinos in Supernovae: New Constraints and Gamma Ray Observables},'' \href{http://dx.doi.org/10.1103/PhysRevD.99.043012}{{\em Phys. Rev.} {\bfseries D99} no.~4, (2019) 043012},
\href{http://arxiv.org/abs/1605.00654}{[{\ttfamily 1605.00654}]}.

\bibitem{2010A&A...522A..16G}
D.~{Gonzalez} and A.~{Reisenegger}, ``{Internal heating of old neutron stars: contrasting different mechanisms},'' \href{http://dx.doi.org/10.1051/0004-6361/201015084}{{\em \aap} {\bfseries 522} (Nov., 2010) A16}, \href{http://arxiv.org/abs/1005.5699}{[{\ttfamily 1005.5699}]}.

\bibitem{1984ApJ...276..325A}
M.~A. {Alpar}, D.~{Pines}, P.~W. {Anderson}, and J.~{Shaham}, ``{Vortex creep and the internal temperature of neutron stars. I - General theory},'' \href{http://dx.doi.org/10.1086/161616}{{\em \apj} {\bfseries 276} (1984) 325--334}.

\bibitem{1984ApJ...278..791A}
M.~A. {Alpar}, P.~W. {Anderson}, D.~{Pines}, and J.~{Shaham}, ``{Vortex creep and the internal temperature of neutron stars. II. VELA pulsar.},'' \href{http://dx.doi.org/10.1086/161849}{{\em \apj} {\bfseries 278} (1984) 791--805}.

\bibitem{Fujiwara:2023tmr}
M.~Fujiwara, K.~Hamaguchi, N.~Nagata, and M.~E. Ramirez-Quezada, ``{Vortex creep heating in neutron stars},'' \href{http://dx.doi.org/10.1088/1475-7516/2024/03/051}{{\em JCAP} {\bfseries 03} (2024) 051}, \href{http://arxiv.org/abs/2308.16066}{[{\ttfamily 2308.16066}]}.

\bibitem{Fujiwara:2023hlj}
M.~Fujiwara, K.~Hamaguchi, N.~Nagata, and M.~E. Ramirez-Quezada, ``{Vortex creep heating vs. dark matter heating in neutron stars},'' \href{http://dx.doi.org/10.1016/j.physletb.2023.138341}{{\em Phys. Lett. B} {\bfseries 848} (2024) 138341}, \href{http://arxiv.org/abs/2309.02633}{[{\ttfamily 2309.02633}]}.

\bibitem{Raj:2024kjq}
N.~Raj, P.~Shivanna, and G.~N. Rachh, ``{Exploring reheated sub-40000 Kelvin neutron stars with JWST, ELT, and TMT},'' \href{http://dx.doi.org/10.1103/PhysRevD.109.123040}{{\em Phys. Rev. D} {\bfseries 109} no.~12, (2024) 123040}, \href{http://arxiv.org/abs/2403.07496}{[{\ttfamily 2403.07496}]}.

\bibitem{1960SvA.....4..187A}
V.~A. {Ambartsumyan} and G.~S. {Saakyan}, ``{The Degenerate Superdense Gas of Elementary Particles},'' {\em \sovast} {\bfseries 4} (Oct., 1960) 187.

\bibitem{Alford:2002rj}
M.~Alford and S.~Reddy, ``{Compact stars with color superconducting quark matter},'' \href{http://dx.doi.org/10.1103/PhysRevD.67.074024}{{\em Phys. Rev. D} {\bfseries 67} (2003) 074024}, \href{http://arxiv.org/abs/nucl-th/0211046}{[{\ttfamily nucl-th/0211046}]}.

\bibitem{PhysRev.95.249}
L.~Van~Hove, ``Correlations in space and time and born approximation scattering in systems of interacting particles,'' \href{http://dx.doi.org/10.1103/PhysRev.95.249}{{\em Phys. Rev.} {\bfseries 95} (Jul, 1954) 249--262}.

\bibitem{Chauhan:2024}
G.~Chauhan, ``{Neutron Stars as probe of Cosmic Neutrino Background}.'' \url{https://indico.sanfordlab.org/event/69/contributions/1468/attachments/889/2202/CosmicNeutrinoBackground_Chauhan.pdf}, 2024.
\newblock Talk at CETUP* 2024.

\bibitem{Chauhan:2024deu}
G.~Chauhan, ``{Neutron Stars as a Probe of Cosmic Neutrino Background},'' \href{http://arxiv.org/abs/2408.01489}{[{\ttfamily 2408.01489}]}.

\bibitem{Leitner:2005}
T.~Leitner, ``{Neutrino Interactions with Nucleons and Nuclei}.'' \url{https://gibuu.hepforge.org/trac/chrome/site/files/dipl/leitner.pdf}, 2005.
\newblock Diplomarbeit.

\bibitem{Budd:2003wb}
H.~S. Budd, A.~Bodek, and J.~Arrington, ``{Modeling quasielastic form-factors for electron and neutrino scattering},'' in {\em {2nd International Workshop on Neutrino-Nucleus Interactions in the Few GeV Region}}.
\newblock 8, 2003.
\newblock \href{http://arxiv.org/abs/hep-ex/0308005}{[{\ttfamily hep-ex/0308005}]}.

\bibitem{Krutov:2002tp}
A.~F. Krutov and V.~E. Troitsky, ``{Extraction of the neutron charge form-factor from the charge form-factor of deuteron},'' \href{http://dx.doi.org/10.1140/epja/i2002-10077-9}{{\em Eur. Phys. J. A} {\bfseries 16} (2003) 285--290}, \href{http://arxiv.org/abs/hep-ph/0202183}{[{\ttfamily hep-ph/0202183}]}.

\bibitem{Garvey:1992cg}
G.~T. Garvey, W.~C. Louis, and D.~H. White, ``{Determination of proton strange form-factors from neutrino p elastic scattering},'' \href{http://dx.doi.org/10.1103/PhysRevC.48.761}{{\em Phys. Rev. C} {\bfseries 48} (1993) 761--765}.

\bibitem{Tomalak:2019ibg}
O.~Tomalak and R.~J. Hill, ``{Theory of elastic neutrino-electron scattering},'' \href{http://dx.doi.org/10.1103/PhysRevD.101.033006}{{\em Phys. Rev. D} {\bfseries 101} no.~3, (2020) 033006}, \href{http://arxiv.org/abs/1907.03379}{[{\ttfamily 1907.03379}]}.

\bibitem{Domcke:2017aqj}
V.~Domcke and M.~Spinrath, ``{Detection prospects for the Cosmic Neutrino Background using laser interferometers},'' \href{http://dx.doi.org/10.1088/1475-7516/2017/06/055}{{\em JCAP} {\bfseries 06} (2017) 055}, \href{http://arxiv.org/abs/1703.08629}{[{\ttfamily 1703.08629}]}.

\bibitem{Shtabovenko:2020gxv}
V.~Shtabovenko, R.~Mertig, and F.~Orellana, ``{FeynCalc 9.3: New features and improvements},'' \href{http://dx.doi.org/10.1016/j.cpc.2020.107478}{{\em Comput. Phys. Commun.} {\bfseries 256} (2020) 107478}, \href{http://arxiv.org/abs/2001.04407}{[{\ttfamily 2001.04407}]}.

\bibitem{Reddy:1997yr}
S.~Reddy, M.~Prakash, and J.~M. Lattimer, ``{Neutrino interactions in hot and dense matter},'' \href{http://dx.doi.org/10.1103/PhysRevD.58.013009}{{\em Phys. Rev. D} {\bfseries 58} (1998) 013009}, \href{http://arxiv.org/abs/astro-ph/9710115}{[{\ttfamily astro-ph/9710115}]}.

\end{thebibliography}\endgroup

\end{document}